\documentclass[fleqn,usenatbib]{mnras}

\usepackage{newtxtext} 
\usepackage[scaled=0.92]{helvet} 

\usepackage[T1]{fontenc}
\usepackage{ae,aecompl}
\usepackage{ulem}

\usepackage{graphicx,epstopdf}	
\usepackage{amsmath}	
\usepackage{amssymb}	
\usepackage{hyperref}

\newcommand{\Msun}{$\mathrm{M}_\odot$}
\newcommand{\Rsun}{$\mathrm{R}_\odot$}
\newcommand{\kms}{km\,s$^{\,-1}$}

\newcommand{\appropto}{\mathrel{\vcenter{
  \offinterlineskip\halign{\hfil$##$\cr
    \propto\cr\noalign{\kern2pt}\sim\cr\noalign{\kern-2pt}}}}}

\newcommand{\blue}[1]{\textcolor{blue}{#1}}

\title[Low-energy Low-mass Fe-CCSNe]
{Low-luminosity type IIP supernovae: SN~2005cs and SN~2020cxd as very low-energy iron core-collapse explosions}

\author[A. Kozyreva et al.]{Alexandra~Kozyreva$^{\,1,2}$\thanks{E-mail: sasha@mpa-garching.mpg.de},
Hans-Thomas Janka$^{\,1}$, Daniel Kresse$^{\,1,3}$, Stefan Taubenberger$^{\,1}$
\newauthor
Petr Baklanov$^{\,4, 5}$
\\
$^{1}$Max-Planck-Institut f\"ur Astrophysik, Karl-Schwarzschild-Str. 1, 85748 Garching, Germany,\\
$^{2}$Alexander von Humboldt Fellow\\
$^{3}$Physik-Department, Technische Universit\"{a}t M\"{u}nchen, James-Franck-Str. 1, 85748 Garching,
Germany\\
$^{4}$NRC ``Kurchatov Institute'' -- ITEP, Moscow, 117218, Russia \\
$^{5}$Keldysh Institute of Applied Mathematics, Russian Academy of Science,
Miusskaya sq. 4, 125047 Moscow, Russia
}
\date{Accepted XXX. Received YYY; in original form ZZZ}

\pubyear{2022}

\begin{document}
\label{firstpage}
\pagerange{\pageref{firstpage}--\pageref{lastpage}}
\maketitle



\begin{abstract}

SN~2020cxd is a representative of the family of low-energy, underluminous
Type~IIP supernovae (SNe), whose observations and analysis were recently 
reported by Yang et al. (2021). Here we re-evaluate the observational data for
the diagnostic SN properties by employing the hydrodynamic explosion model
of a 9\,M$_\odot${} red supergiant progenitor with an iron core and a pre-collapse
mass of 8.75\,M$_\odot${}. The explosion of the star was obtained by the
neutrino-driven mechanism in a fully self-consistent simulation in three
dimensions (3D). Multi-band light curves and photospheric velocities for the
plateau phase are computed with the one-dimensional radiation-hydrodynamics
code \verb|STELLA|, applied to the spherically averaged 3D explosion model as well
as sphericized radial profiles in different directions of the 3D model.
We find that the overall evolution of the bolometric light curve, duration
of the plateau phase, and basic properties of the multi-band emission
can be well reproduced by our SN model with its explosion energy of only
$0.7\times 10^{50}$\,erg and an ejecta mass of 7.4\,M$_\odot${}. These values
are considerably lower than the previously reported numbers, but they are
compatible with those needed to explain the fundamental observational 
properties of the prototype low-luminosity SN~2005cs. Because of the good
compatibility of our photospheric velocities with line velocities determined
for SN~2005cs, we conclude that the line velocities of SN~2020cxd
are probably overestimated by up to a factor of about 3. The evolution of
the line velocities of SN~2005cs compared to photospheric velocities in
different explosion directions might point to intrinsic asymmetries in
the SN ejecta.
\end{abstract}





\section[Introduction]{Introduction}
\label{sect:intro}

The recently discovered supernova (SN) 2020cxd is a low-luminosity
hydrogen-rich type~II SN \citep[LL type IIP SN,][]{2021AandA...655A..90Y,2022arXiv220303988V}. 
Due to its low plateau luminosity of about $10^{\,41}$~erg\,s$^{\,-1}${} 
and relatively low photospheric velocity at the end of the
plateau,
it is considered to be a member of the family of LL type IIP SNe
\citep[e.g., ][]{2009MNRAS.394.2266P,2014MNRAS.439.2873S,2021MNRAS.501.1059R}. 
Analysing the bolometric light curve, \citet{2021AandA...655A..90Y} 
inferred the explosion of a red supergiant (187~\Rsun{}) with a pre-collapse 
mass of about 11~\Msun{} (and a zero-age main sequence (ZAMS) mass of about 12~\Msun{}) 
and an energy of 0.58~foe\footnote{The authors report a kinetic energy of
0.43~foe and a thermal energy of 0.15~foe, which results in a total energy of
0.58~foe.} (1~foe\,$=10^{\,51}$~erg). SN~2020cxd exhibits a large drop between
the plateau and the radioactive tail, as well as a very low tail luminosity. The 
total mass of the ejected radioactive nickel {}$^{56}$Ni{} is 0.003~\Msun{}. More recently, \citet{2022arXiv220303988V} found that the best-fit SN parameters are an ejecta mass of 7.5~\Msun{}, explosion energy of 0.097~foe, and a progenitor radius of 575~\Rsun{}. These values are in
fairly satisfactory agreement with the results we will report in our paper.

We note in passing that 
besides their relevance for stellar and SN astrophysics, LL type IIP SNe are also of great 
interest for setting constraints to physics beyond the Standard Model (BSM) of particle physics
\citep{Caputo+2022}. BSM particle production might play a role at the high densities and temperatures in new-born neutron stars. This could extract energy from the compact remnant during the neutrino-cooling phase and would thus change the observable neutrino signal \citep[e.g.,][]{2018JHEP...09..051C,2019JCAP...10..016C,Caputo+2022a}, and it could also have an impact on the SN energetics,
dynamics, and electromagnetic emission in various ways \citep[e.g.,][]{Rembiasz+2018,Mori+2022,Caputo+2022}.

According to the current understanding, based on recent self-consistent
3-dimensional (3D) simulations of the underlying mechanism that
causes the explosion of massive stars, SNe with a plateau in
their light curves result from the neutrino-driven mechanism 
\citep[e.g.,][]{2014ApJ...786...83T,2015ApJ...807L..31L,2015ApJ...801L..24M,2015ApJ...808L..42M,2017MNRAS.472..491M,2018MNRAS.479.3675M,2018ApJ...855L...3O,2018ApJ...852...28S,2019MNRAS.485.3153B,2019ApJ...873...45G,2020MNRAS.496.2039S,2021ApJ...915...28B}. The amount of radioactive nickel {}$^{56}$Ni{}
produced in this kind of explosion is proportional to the neutrino luminosity and
amount of ejecta material heated by neutrinos. The higher the neutron-star mass
(correlated with the progenitor's compactness as defined by \citealt{2011ApJ...730...70O})
the higher the neutrino luminosity, 
the bigger the amount of neutrino-heated ejecta, the more energetic the 
shock wave, and the larger the mass of radioactive nickel
\citep{2015PASJ...67..107N,2016ApJ...818..124E,2016MNRAS.460..742M,2016ApJ...821...38S,2017hsn..book.1095J,2020ApJ...890...51E}.
These dependencies associated with neutrino-driven explosions are in line
with correlations between SN energies, {}$^{56}$Ni{} masses and plateau
luminosities deduced from observations \citep{2015ApJ...799..215P,2017ApJ...841..127M,2020rfma.book..189P}.
Recent studies show that the reference amount of ejected radioactive nickel {}$^{56}$Ni{} is
about 0.03~\Msun{} for a diagnostic explosion energy around 0.6~foe 
\citep{2016MNRAS.460..742M,2017MNRAS.472..491M,2020ApJ...890...51E}.
Therefore, the parameters of SN~2020cxd diagnosed by \citet{2021AandA...655A..90Y}
are in conflict with 
our current theoretical and observational picture of the underlying physics of core-collapse SN explosions.
Motivated by this fact we present here a revision of the analysis made by
\citet{2021AandA...655A..90Y} with the effort to
find a more reliable interpretation for the progenitor of SN~2020cxd and its
explosion.
We also choose SN~2005cs as a representative member of
the family of LL type IIP SNe, because SN~2005cs has a better data coverage,
and analyse SN~2020cxd together with SN~2005cs in the context of our theoretical model.

In Section~\ref{sect:method} we describe our considered SN model and the
radiation-hydrodynamics treatment with the \verb|STELLA| code.  In
Section~\ref{sect:results} we
present bolometric and multi-band light curves as well as photospheric
velocities and colours of our model calculations in comparison to observational data for
SN~2020cxd and SN~2005cs, including an assessment of the line velocities
reported for SN~2020cxd in the literature.  In Section~\ref{sect:conclusions} we summarize our
results and draw conclusions.  Appendices~\ref{appendix:append1},
\ref{appendix:append2}, and \ref{appendix:append3} present angle-averaged and
direction dependent profiles for our considered 3D SN model as well as
corresponding multi-band light curves and the colour evolution of the SN
calculations.


\section[Input models]{Input model and Method}
\label{sect:method}

\begin{table*}
\caption{Parameters of the LL type IIP SNe 2020cxd, 2005cs, and the
angle-averaged parameters of model s9.0. $t_\mathrm{p}$ is the duration
of plateau phase which ends at the middle of the transition to the
radioactive tail.
Velocity gives the H$\alpha${} line velocity for SN~2020cxd and the Sc\,\textsc{ii} line velocity for 
SN~2005cs around the middle of the plateau, radius is the progenitor radius at the pre-collapse stage,
$M_\mathrm{ej}$ the ejecta mass, $M_\mathrm{prog}$ the progenitor mass, $M_{^{56}\mathrm{Ni}}$
the ejected mass of $^{56}$Ni, and $E_\mathrm{expl}$ the explosion energy (for model s9.0
to be identified with $E_\mathrm{kin}$ in Table~\ref{table:rays}). }
\label{table:data}
\begin{tabular}{|l|c|c|c|c|c|c|c|c|l}
\hline
SN & $M_\mathrm{bol}$  & $t_\mathrm{p}$ & Velocity& Radius  & $M_\mathrm{ej}$& $M_\mathrm{prog}$ & $M_{^{56}\mathrm{Ni}}$ &$E_\mathrm{expl}$ & Reference\\
   & [mags]& [days]& [1000\,km\,s$^{\,-1}$]&[\Rsun] & [\Msun]        & [\Msun]           & [\Msun] & foe & \\
\hline
2020cxd&$-$14.0&130& 4& 187 & 9.5            &          & 0.003 & 0.58 & \citet{2021AandA...655A..90Y}\\
 & & & & 575 & 7.5 & & 0.0018 & 0.097 & \citet{2022arXiv220303988V}\\ 
2005cs &$-$14.58&130&1.5& 100 & 11.1           &          & 0.0028& 0.3  & \citet{2009MNRAS.394.2266P}\\
       &&&& 357 & 9.5            &          & 0.006& 0.16 & \citet{2017MNRAS.464.3013P}\\
       &&&& 600 & 15.9           & 18.2     & 0.0082& 0.41 & \citet{2008AandA...491..507U}\\
\hline
s9.0   &&&& 408 & 7.4 & 8.75 & 0.003  & 0.068 & present study\\
\hline
\end{tabular}
\end{table*}

In Table~\ref{table:data}, we report the published properties of SN~2005cs as a representative member of
the family of LL type IIP SNe and of the low-luminosity SN~2020cxd, and 
confront them with the properties of our theoretical model.
We consider a star near the low-mass end of SN progenitors with an initial mass 
of 9~\Msun{} \citep{2016ApJ...821...38S} as the most suitable model.
A low-energy neutrino-driven explosion of such a low-mass progenitor
can explain LL type IIP SNe, as was
already
discussed by a number of observational and
theoretical studies
\citep{2008ApJ...688L..91M,2012AJ....143...19V,2018MNRAS.475..277J}. We note
that the study by \citet{2008AandA...491..507U} explains SN~2005cs
with an explosion of a higher-mass progenitor of 18.2~\Msun{}, however,
their study was based on a non-evolutionary progenitor configuration. A
connection with the explosion of such a high-mass progenitor, however, is
disfavoured by recent self-consistent 3D models of neutrino-driven explosions of
18\,--\,19~\Msun{} stars \citep{2017MNRAS.472..491M,2021ApJ...915...28B}. 

\begin{figure*}
\centering
\includegraphics[width=0.5\textwidth]{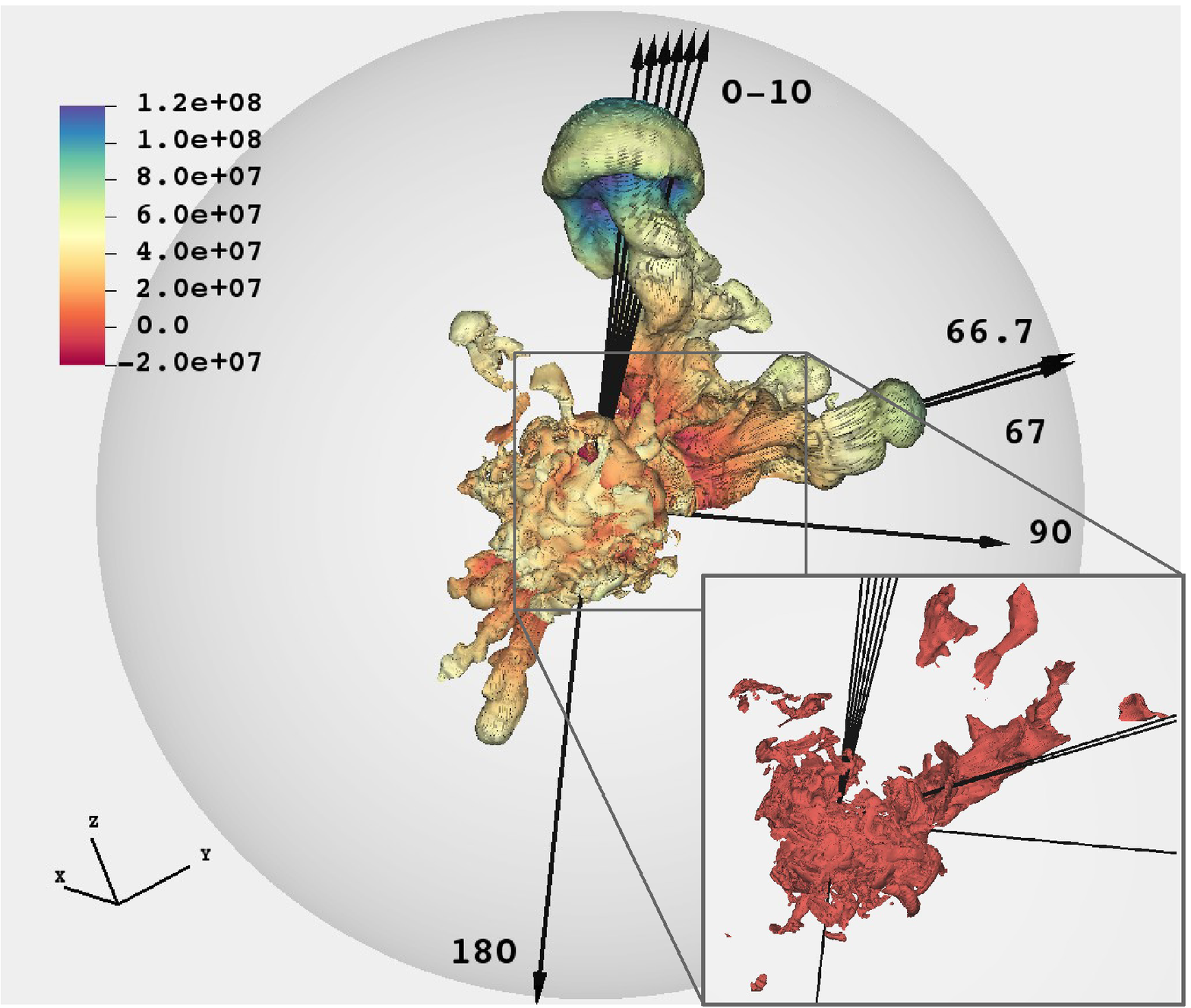}~
\includegraphics[width=0.5\textwidth]{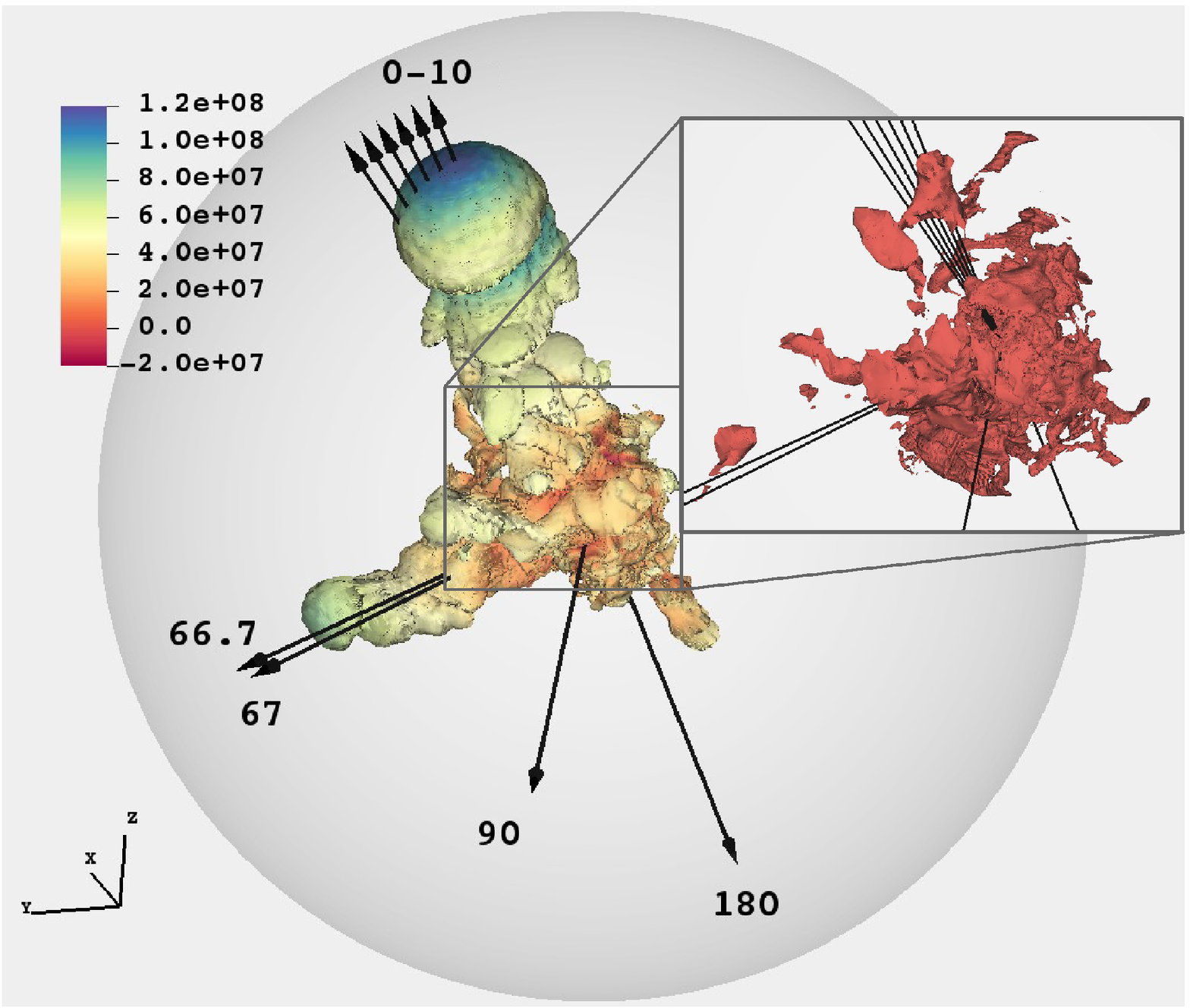}
\caption{Three-dimensional explosion geometry of the s9.0 model
\citep{2020MNRAS.496.2039S} used as input for the present study.  The
two panels show different views of the isosurface of a constant
{}$^{56}$Ni{} mass fraction 
of 0.01 at 1.974 days after core collapse ($\sim$0.2 days before the
fastest part of the deformed SN shock reaches the stellar surface), with the
radial velocity (in units of cm\,s$^{\,-1}$) colour-coded.  The zoom insets show
the isosurface of a constant {}$^{56}$Ni{} mass fraction of 0.1 with the same
colour-coding.  The
semi-transparent grey sphere marks the stellar surface (at a radius of
$2.86\times10^{\,13}$\,cm).  Black arrows denote the selected angular
directions as listed in Table~\ref{table:rays}; for reasons of clarity we do not
show the ``$-$'' directions (i.e., $-2^\circ$ to $-10^\circ$), which are distinguished from the ``$+$'' directions ($+2^\circ$ to $+10^\circ$) by an azimuthal angular shift of 180$^\circ$ around the major {}$^{56}$Ni{} plume axis (i.e., the 0$^\circ$ direction).}
\label{figure:asymmetry}
\end{figure*}

Our progenitor model of a star with an initial mass of 9~\Msun{} (model s9.0 of \citealt{2016ApJ...821...38S})
was computed with the stellar evolution code \verb|KEPLER|
\citep{1978ApJ...225.1021W}
until the onset of iron-core collapse \citep{2015ApJ...810...34W}.
A one-dimensional (1D) SN simulation of this model, which was
exploded parametrically via neutrino heating by \citet{2016ApJ...818..124E} and
\citet{2016ApJ...821...38S}, led to an explosion energy of 0.11 foe.  This
explosion model was later used for nebular spectral modeling by
\citet{2018MNRAS.475..277J}.
The corresponding self-consistent 3D simulation of a neutrino-driven explosion of this 9\,M$_\odot${}
progenitor model with the \verb|PROMETHEUS-VERTEX| code was first
discussed in \citet{2020ApJ...891...27M} and was later used for long-time 3D simulations with 
\verb|PROMETHEUS-HOTB| by
\citet{2020MNRAS.496.2039S}. Independently, the s9.0 model was successfully exploded in 3D by
\citet{2019ApJ...873...45G} and by \citet{2019MNRAS.485.3153B}.

For our present work we use the s9.0 model of \citet{2020MNRAS.496.2039S}, which has already been used as input for the study by \citet{2021MNRAS.503..797K}.
We note that an improved post-processing analysis is applied in the present work and the profiles
are extracted from the 3D simulation at an earlier epoch, namely at 1.974~days (instead of 2.823~days)
after core collapse. At this moment the fastest parts of the
shock are within 0.2~days before their breakout from the stellar surface.
The shock front is strongly deformed, and the slowest parts of the shock
reach the stellar surface only approximately 1~day later, i.e. about 3.1~days after the onset of the explosion.
In Appendix~\ref{appendix:append1}, we present comparative plots showing the structure of the angle-averaged profiles used in \citet{2021MNRAS.503..797K} compared to the profiles used as input in the present work.
The geometry of the shock front is far from
spherical symmetry and the distribution of the ejected {}$^{56}$Ni{} displays a main plume and a few smaller plumes of
high-entropy material. We illustrate the asphericity of the model in
Figure~\ref{figure:asymmetry}. In total we extract 16 radial profiles
which represent the structure of the ejecta in different angular directions
of the 3D model.
The profiles are listed in Table~\ref{table:rays} and 
are named according to the angle relative to the central-axis direction of the
fastest plume of the ejected {}$^{56}$Ni{} (see Figure~{\ref{figure:asymmetry}}). 
The velocity and density profiles along with the distribution of the 
{}$^{56}$Ni{} mass fraction in each of the different radial directions are 
shown in Appendix~\ref{appendix:append2}.
We note that the value of the explosion energy of 0.048~foe
as given in Table~4 of \citet{2020MNRAS.496.2039S} is taken at the
end of the \verb|PROMETHEUS-VERTEX| simulation at 3.14~s, whereas
we give the value at 1.974~days from the long-time continuation run with
the \verb|PROMETHEUS-HOTB| code.

In the current study, the 3D explosion model is mapped into the 1D radiation-hydrodynamics
code \verb|STELLA| \citep{2006A&A...453..229B}.  \verb|STELLA| is
capable of processing hydrodynamics as well as the radiation field
evolution, i.e. computing light curves, spectral energy distribution and resulting
broad-band magnitudes and colours.  We use the standard parameter settings,
well-explained in many papers involving \verb|STELLA| simulations \citep[see
e.g.,][]{2019MNRAS.483.1211K,2020MNRAS.497.1619M}.  The
thermalisation parameter, which accounts for the treatment of the line 
opacity and expresses the ratio between absorption and scattering opacities, 
is set to 0.9 as recommended by the recent study by \citet{2020MNRAS.499.4312K}.
The profiles along different radial directions are sphericized during the mapping
into \verb|STELLA|. 
The corresponding 4$\pi$-equivalent values of different quantities are listed in 
Table~\ref{table:rays}. The total amount
of {}$^{56}$Ni{} is scaled to 0.003~\Msun{} to match the mass of {}$^{56}$Ni{}
estimated for SN~2020cxd \citep{2021AandA...655A..90Y}. 
A value of the {}$^{56}$Ni{} (plus {}$^{56}$Co{} and {}$^{56}$Fe{}) yield 
of 0.00635~\Msun{} was reported by
\citet{2020MNRAS.496.2039S} for the 3D explosion model of s9.0 and recently determined to be 0.0057~\Msun{} by 
a more accurate re-evaluation of the simulation outputs (see data in
\href{https://wwwmpa.mpa-garching.mpg.de/ccsnarchive/data/Stockinger2020/}{https://wwwmpa.mpa-garching.mpg.de/ccsnarchive/data/Stockinger2020/}). 
Note that the angle-averaged
profile in \citet{2021MNRAS.503..797K} had 0.005~\Msun{} of {}$^{56}$Ni{}.
We mention two reasons why a rescaling of the $^{56}$Ni mass is well 
motivated:
\begin{enumerate}
\item The nucleosynthesis calculations yielding the values reported by
\citet{2020MNRAS.496.2039S} and re-evaluated in the present study 
were only approximate, because they were based on the use of a small nuclear
$\alpha$-chain network.
\item The exact amount of {}$^{56}$Ni{} produced in 3D explosion models is
extremely sensitive to the electron fraction ($Y_e$) in the neutrino-heated
ejecta, which is determined by the interaction of the innermost ejecta with
the intense neutrino radiation from the proto-neutron star.  Therefore, the
$Y_e$ distribution in these ejecta depends sensitively on the details of the
neutrino physics and neutrino transport, including the still incompletely
understood effects of neutrino-flavor oscillations in and near the proto-neutron
star.
\end{enumerate}
The uncertainties in the exact ejected mass of {}$^{56}$Ni{} corresponding
to these points can well amount to a factor of 2.
 


\begin{table*}
\caption{Isotropic-equivalent parameter values for the profiles corresponding to different radial directions
and for the angle-averaged profile (labeled ``AVG'') taken from the 3D model s9.0 \citep{2020MNRAS.496.2039S}. The name of the direction represents the
angle from the axis of the main high-entropy plume.
For the cases of $2^\circ$--$10^\circ$ we show two (``$+$'' and ``$-$'') directions,
which are distinguished by an azimuthal angular shift of 180$^\circ$ around the axis of the major {}$^{56}$Ni{} plume (i.e., the $0^\circ$ direction).
The total mass of radioactive
{}$^{56}$Ni{} is a sum of {}$^{56}$Ni{} and its daughter products
{}$^{56}$Co{} and {}$^{56}$Fe{}. Note that we list the {}$^{56}$Ni{} mass
($M_\mathrm{^{56}Ni}$)
with the value that is present in each of the profiles before scaling to 0.003~\Msun{} 
needed to match the mass of {}$^{56}$Ni{} in SN~2020cxd and SN~2005cs. 
$M_\mathrm{tot}$ is the isotropic-equivalent total mass including the central compact object, while $M_\mathrm{ej}$ is the isotropic equivalent of the ejected mass. $E_\mathrm{expl}$ is the $4\pi$-equivalent explosion energy at 1.974 days.
$E_\mathrm{kin}$ is the analogously determined terminal kinetic energy at the end of
the radiative transfer simulations (at day~170), before and after 
scaling the {}$^{56}$Ni{} mass to 0.003\,M$_\odot${}.
The energy difference is a consequence of the different release of energy in
radioactive {}$^{56}$Ni{} and {}$^{56}$Co{} decay which 
contributes to the energy balance
after the energies of $\gamma$-rays and positrons are thermalised.}
\label{table:rays}
\begin{tabular}{|r|c|c|c|c|c|c|}
\hline
direction& $M_\mathrm{^{56}Ni}$ & $M_\mathrm{tot}$ & $M_\mathrm{ej}$ & $E_\mathrm{expl}$ & $E/M$ & $E_\mathrm{kin}$ \\
     & [M$_\odot${}] & [M$_\odot${}]   & [M$_\odot${}]  & [0.1~foe] & [foe/M$_\odot${}] & [0.1~foe] \\
     
\hline
0$^\circ$    &  0.0974 & 5.75 & 4.40 & 1.2812 & 0.0303 & 1.3337/1.2641\\
$+$2$^\circ$ &  0.0886 & 5.96 & 4.61 & 1.3069 & 0.0293 & 1.3502/1.2915\\
$-$2$^\circ$ &  0.0951 & 5.48 & 4.13 & 1.1885 & 0.0299 & 1.2362/1.1701\\
$+$4$^\circ$   &  0.1229 & 6.14 & 4.79 & 1.2964 & 0.0286 & 1.3684/1.2801\\
$-$4$^\circ$   &  0.1150 & 5.70 & 4.35 & 1.1553 & 0.0278 & 1.2110/1.1289\\
$+$6$^\circ$   &  0.1278 & 6.23 & 4.88 & 1.1894 & 0.0259 & 1.2626/1.1641\\
$-$6$^\circ$   &  0.0947 & 5.79 & 4.44 & 1.0850 & 0.0253 & 1.1233/1.0543\\
$+$8$^\circ$   &  0.0947 & 6.38 & 5.03 & 1.0524 & 0.0217 & 1.0931/1.0177\\
$-$8$^\circ$   &  0.0540 & 5.85 & 4.50 & 0.9425 & 0.0210 & 0.9459/0.9070\\
$+$10$^\circ$  &  0.0814 & 6.50 & 5.15 & 0.8836 & 0.0177 & 0.9122/0.8445\\
$-$10$^\circ$  &  0.0460 & 6.12 & 4.77 & 0.7960 & 0.0166 & 0.7930/0.7583\\
66.7$^\circ$ &  0.0659 & 8.68& 7.32& 0.8026 & 0.0117 & 0.8594/0.7951\\
67$^\circ$   &  0.0582 & 8.66 & 7.31 & 0.8030 & 0.0117 & 0.8523/0.7964\\
90$^\circ$   &  0.0010& 8.75 & 7.40 & 0.7596 & 0.0102 & 0.7524/0.7544\\
180$^\circ$  &  0.0026 & 8.80 & 7.45 & 0.6943 & 0.0092 & 0.6884/0.6887\\
\hline
AVG & 0.0057 & 8.75 & 7.40 & 0.7334 & 0.0095& 0.6790/0.6765\\
\hline
\end{tabular}
\end{table*}

Recently, radiative transfer simulations with the spherically averaged 3D 
explosion model of s9.0 were done in a separate paper and compared to the
LL type IIP SNe 2005cs and 1999br \citep{2021MNRAS.503..797K}. The
model matches the broad-band magnitudes and bolometric light curves of these
observed SNe.
The comparison demonstrates the applicability of the given model for
LL type IIP SNe. Hence, it has already been shown that our explosion model of
the 9~\Msun{} progenitor is
capable of explaining a number of members of the family of LL type IIP SNe.


\section[Results]{Results}
\label{sect:results}

\subsection[Bolometric properties]{Bolometric properties}
\label{subsect:bol}

\begin{figure}
\includegraphics[width=0.5\textwidth]{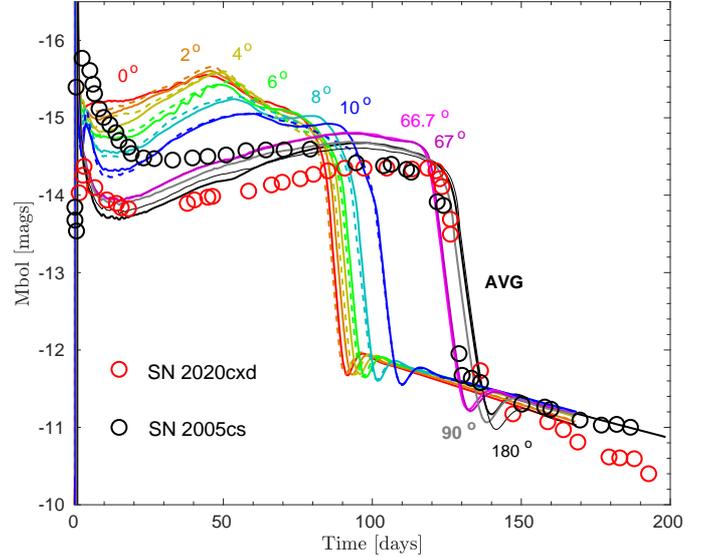}
\caption{Bolometric light curves (in absolute magnitudes) for different 
radial directions of model s9.0, 
and LL type IIP SNe~2005cs and 2020cxd. The label ``AVG'' represents the light curve for the angle-averaged profile. The solid and dashed curves indicate the ``$+$'' and ``$-$'' directions for the cases of 2$^\circ$--10$^\circ$, as introduced in Table~\ref{table:rays}.}
\label{figure:bol}
\end{figure}

In Figure~\ref{figure:bol} bolometric light curves 
for different radial directions of our 3D explosion simulation and the angle-averaged
case of model s9.0 are displayed.
We superpose bolometric light curves of two LL type IIP SNe, SN~2020cxd and
SN~2005cs.\footnote{It should be kept in mind that the observational data represent approximations to the bolometric light curves (either constructed from the near-UV to near-IR bands or from a dilute-blackbody fit to the optical bands), whereas the theoretical data are true bolometric results.} 
Our 66.7$^\circ${}, 67$^\circ${},
90$^\circ${}, 180$^\circ${}, and angle-averaged cases are capable of
matching the bolometric light curve of SN~2020cxd to a very large extent without
any artificial tuning besides a proper scaling of the {}$^{56}$Ni{} mass to
match the observationally determined value (see Section~\ref{sect:method}). And 
they are also able to reproduce the global behaviour of the 
bolometric light curve of SN~2005cs.

Because the explosion energy is a crucial
factor governing the plateau luminosity
\citep{1980ApJ...237..541A,1993ApJ...414..712P,2009ApJ...703.2205K,2016ApJ...821...38S,2019ApJ...879....3G,2019MNRAS.483.1211K},
a luminosity of a Type~IIP SN as low as 
$L \simeq
10^{\,41}$~erg\,s$^{\,-1}$ or --14~mags (average for the LL type IIP family) can be
reproduced by very low explosion energies, around or below 0.1~foe for
stars near the low-mass end of core-collapse SN progenitors \citep{2017MNRAS.464.3013P}. 
The \emph{V}-band magnitude ($V$) and the bolometric luminosity ($L$) at the middle of
the plateau are:
\begin{equation}
\begin{array}{l}
V \sim 1.25 \log M - 2.08 \log E - 1.67 \log R\,,       \\  
\log L \sim - 0.4 \log M + 0.74 \log E + 0.76 \log R\,,     
\end{array}
\label{equation:PopovLbol}
\end{equation}
where $M$ is the ejecta mass, $E$ the explosion energy, and $R$ the
progenitor radius prior to the explosion. The scaling relations are taken from
\citet{1993ApJ...414..712P} and \citet{2019ApJ...879....3G}, respectively. 
Note that we neglect the additive terms given in these references, because we are only 
interested in the dependencies. If the energy varies between 0.03~foe and 1.5~foe
\citep{2016ApJ...818..124E,2020MNRAS.496.2039S}, this corresponds to $-$1.5 and 0.2 in logarithmic
scale, which results in a luminosity scatter up to 1~dex.
The low explosion energy required for low-luminosity events can be released in low-mass progenitors
of 9--10~\Msun{}, as the energy exhibits a (rough) tendency to increase with the initial progenitor mass
\citep{2015PASJ...67..107N,2016ApJ...818..124E,2016MNRAS.460..742M,2020rfma.book..189P}.
The second strongest factor is the progenitor radius, and the weakest effect is connected to the ejecta mass.
The radius ranges between 100~\Rsun{} and a few 1000~\Rsun{} for the observed red-supergiants
\citep{2005ApJ...628..973L,2006ApJ...645.1102L}. This leads to a scatter
in luminosity of 0.76~dex. The ejecta mass tends to be around 10~\Msun{} and
has a relatively less important effect.
To reach very low luminosities, the radius of the progenitors should not be very
large, yet it should still allow for sufficient
recombination to power the light curve \citep{1976Ap&SS..44..409G} 
and to provide an extended
plateau phase, since shrinking the progenitor radius shortens the duration of the plateau:
\begin{equation}
\begin{array}{l}
\log t_\mathrm{p} \sim 0.5 \log M - 0.167 \log E + 0.167 \log R\,,\\
\log t_\mathrm{p} \sim 0.41 \log M - 0.28 \log E - 0.02 \log R\,+0.13 \log M_\mathrm{Ni}\,,
\end{array}
\label{equation:PopovTpl}
\end{equation}
where $t_\mathrm{p}$ is the plateau duration, $M$, $E$, and $R$ have the same meaning
as in Equation~\ref{equation:PopovLbol}, and $M_\mathrm{Ni}$ is the total mass
of radioactive {}$^{56}$Ni{}. The relations are taken from
\citet{1993ApJ...414..712P} and \citet{2019ApJ...879....3G}, respectively. Note that the updated scaling relation for the plateau duration from \citet{2019ApJ...879....3G} does not show a strong dependence on the progenitor radius.
Even though the highest exponent is in the mass term, the mass does not strongly affect the
variations of the duration since the ejecta mass is always close to 10~\Msun{}.
The influence of radioactive nickel in the ejecta on the
plateau luminosity is moderate, whereas the plateau duration is strongly
affected by the presence of nickel and its distribution \citep{2019MNRAS.483.1211K}.
However, for low-mass explosions, which we consider as the most probable
explanation for the LL type IIP SNe, the {}$^{56}$Ni{} yield is a few
thousandths of a solar mass and does not extend the length of the plateau and its 
luminosity value to any significant extent \citep[see][]{2021MNRAS.503..797K}.

The light curves for the directions aligned with the fastest parts of the s9.0
ejecta, i.e. close to the directions of {}$0^\circ - 10^\circ${},
do not match the given observed LL type IIP SNe. Nevertheless, we presume
that flux from these angles may contribute to other directions.
To clarify the question on the contributions of other radial directions to the particular
viewing direction, realistic 3D radiative-transfer simulations are required, which may be
carried out in the future. 

Interestingly, the bolometric light curve during the plateau increases for SN~2020cxd until shortly before
the steep decline to the nickel tail, which is amazingly well reproduced by our explosion model for the 
angle-averaged profiles and the directions at 66.7$^\circ${}, 67$^\circ${}, 90$^\circ${}, and 180$^\circ${}. 
In contrast, SN~2005cs shows nearly a flat plateau and after the middle of the plateau a shallow decline
sets in well before the steep drop to the radioactive tail. 
Such a difference in the shape of the plateau was discussed in \citet{2019MNRAS.483.1211K}.
It could, in principle, be connected to different degrees of mixing of radioactive
{}$^{56}$Ni{}. In the context of the considered LL type IIP SNe, however, this explanation
is disfavored by the small mass of only 0.003\,M$_\odot$ of ejected $^{56}$Ni \citep{2021MNRAS.503..797K}.


The different light-curve shapes
of SN~2005cs and SN~2020cxd might instead be a hint of a considerable degree of explosion asymmetry in the 
former case, even already during the shock-breakout phase and during the early expansion of the hydrogen envelope.
Such a possibility for SNe of low-mass progenitors is suggested by the extreme deformation
of the ejecta associated with the largest nickel plume in our 3D explosion model s9.0. This
nickel-rich plume extends through the entire hydrogen envelope and pushes the expansion of 
the SN shock wave. Therefore, when the head of the plume reaches the stellar surface, the
shock breaks out roughly 1\,day earlier than in the other directions (see 
\citealt{2020MNRAS.496.2039S} and our discussion in Section~\ref{sect:method} connected to 
Figure~\ref{figure:asymmetry}). The light curves computed for sphericized ejecta conditions
in different directions of our 3D explosion model (as displayed in Figure~\ref{figure:bol})
suggest the possible influence of such explosion asymmetries. In the directions from 0$^\circ$ 
to 10$^\circ$ relative to the axis of the biggest nickel-rich plume, the higher explosion energy
per solid angle leads to faster expansion of the ejecta and of the photospheric radius and thus 
lifts the light curve during the initial decline to the plateau and the early phase of the 
plateau. This might account for the higher luminosity of SN~2005cs during the first 20~days 
and the flat evolution of its plateau afterwards until about 80--90~days. The luminosity 
during this phase could therefore be boosted by the emission from faster ejecta for an 
explosion with considerable asymmetry in the outer hydrogen envelope, whereas SN~2020cxd 
would not be influenced by such effects, at least not for our viewing direction. At later 
times during the plateau phase, when the emitted radiation escapes from deeper layers,
where the radioactive nickel and ejecta energy are more spherically distributed 
(see Figure~\ref{figure:asymmetry}), the outer asymmetries play no important role any
longer and the light curves become more similar, as seen for SN~2005cs and SN~2020cxd
after about 90~days (Figure~\ref{figure:bol}). Of course, based on our 1D radiative
transfer modeling such an interpretation remains speculative. It requires confirmation
by 3D light-curve calculations for different viewing angles for our 3D explosion model
s9.0 or other low-mass 3D SN models with large-scale explosion asymmetries.

\begin{figure}
\centering
\vspace{3mm}
\includegraphics[width=0.5\textwidth]{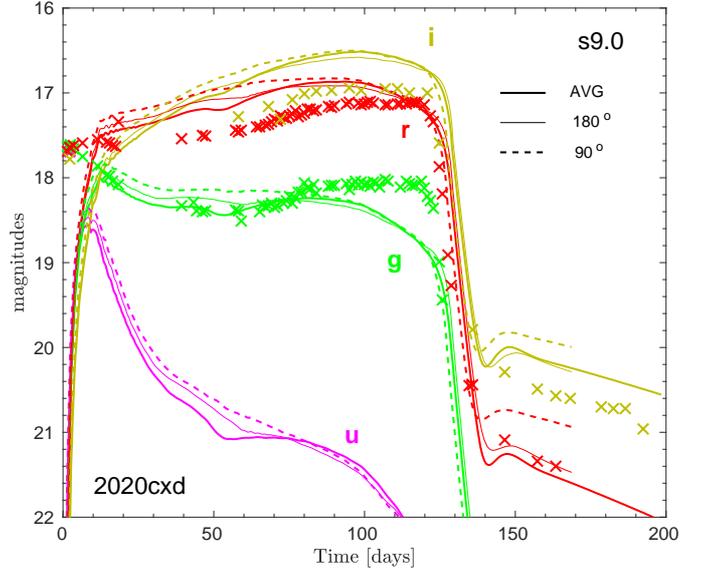}
\caption{$ugri$ broad-band light curves (in apparent magnitudes) 
for selected directions of model s9.0 and SN~2020cxd \citep{2021AandA...655A..90Y}.}
\label{figure:bands2020cxd}
\end{figure}

\begin{figure}
\vspace{3mm}
\centering
\includegraphics[width=0.5\textwidth]{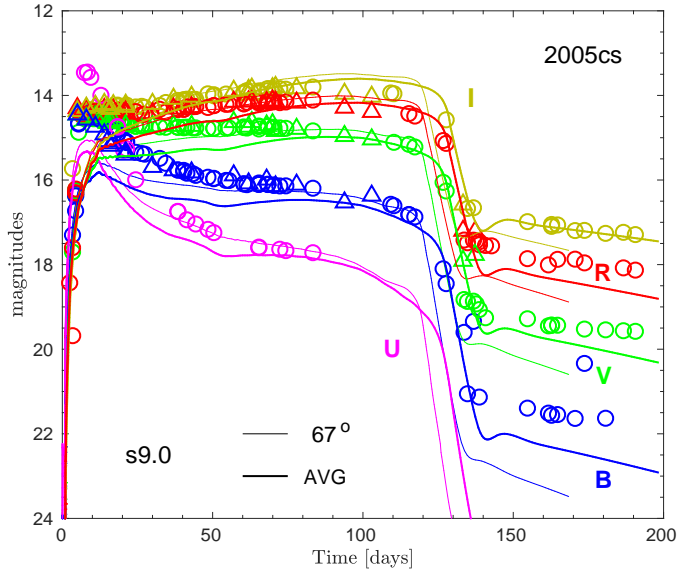}
\caption{$U\!BV\!RI$ broad-band light curves (in apparent magnitudes) for selected
directions of model s9.0 and SN~2005cs. The observational data for SN~2005cs are taken from  \citet{2006AandA...460..769T} (triangles) and from \citet{2009MNRAS.394.2266P} (circles).}
\label{figure:bands2005cs}
\end{figure}

\subsection[Broad-band light curves]{Broad-band light curves compared
to LL type IIP SNe}
\label{subsect:bands}

We present $U\!BV\!R$ broad-band light curves for all considered angular directions 
of our model s9.0 in Appendix~\ref{appendix:append3}. In this section we show 
the results for a few selected angular directions suitable
to reproduce the light curve properties of LL type IIP SNe 2020cxd and 2005cs
(Figure~\ref{figure:bands2020cxd} and Figure~\ref{figure:bands2005cs}, respectively).
In Figure~\ref{figure:bands2020cxd}, $ugri$ magnitudes for the 90$^\circ${}, 180$^\circ${}, and
angle-averaged cases are shown together with those observed in SN~2020cxd. It is
difficult to draw conclusions about the relevance of the synthetic light curves
because of the limited set of filters used for the photometry of SN~2020cxd.
However, the general behaviour of the fluxes in broad bands is explained by
our selected cases. Specifically, the flux in the $g$-band is reproduced
by the directions outside the main plume of s9.0 or the angle-averaged case to good agreement.

At the same time, the 66.7$^\circ${}, 67$^\circ${},
90$^\circ${}, 180$^\circ${}, and angle-averaged models, match the general behaviour of the broad-band
magnitudes of SN~2005cs \citep{2006MNRAS.370.1752P,2006AandA...460..769T,2009MNRAS.394.2266P}, 
except during the first $\sim$50\,days when the models
underestimate the flux in all bands. However, the same is true for the
comparison with the bolometric light curve of SN~2005cs as discussed in
Section~\ref{subsect:bol}. We show only light
curves for the 67$^\circ${} and angle-averaged cases in Figure~\ref{figure:bands2005cs}, 
since the rest of the suitable cases range between the 67$^\circ${} and angle-averaged curves.
We note that the results presented in this figure are from the new radiative-transfer simulations
carried out on the basis of our updated profiles extracted from the 3D SN simulation of s9.0.
There are only minor differences compared to the multi-band light curves published in 
\citet{2021MNRAS.503..797K} for the previously used initial profiles of this explosion model, 
but we include the figure here also to testify these slight differences. In Figure~\ref{figure:colours} (Appendix~\ref{appendix:append3}), we additionally present the $B\!-\!V$ and $V\!-\!R$ colours for all 
considered radial directions of our 3D model s9.0 and compare them to the observed colours of SN~2005cs. 
The colour analysis shows that the overall trends of these observables of SN~2005cs are satisfactorily 
reproduced by the colour evolution of our model. During the first 50~days the rise of $V\!-\!R$ is 
somewhat better followed by the radial directions within the biggest plume ($0^\circ-10^\circ$), whereas the
subsequent increase of $V\!-\!R$ at later times is close to that for the angle-averaged model and the radial 
directions outside of the main plume. In the case of $B\!-\!V$ the observational data are bracketed by the 
angle-averaged model and the large-angle directions on the one side and those for $0^\circ-10^\circ$ on
the other side until about day 35. Subsequently, the data follow closely the evolution described by the
angle-averaged model and the large-angle directions.

Nevertheless, we conclude that our angle-averaged model of s9.0 and the directions
outside the main plume explain bolometric and broad-band light curves of the 
considered observational examples of the LL type IIP SN family sufficiently well.
This is particularly noteworthy because we do not tune the
output of the self-consistent 3D explosion simulations while
mapping it into the radiative-transfer code \verb|STELLA| except for adjusting the {}$^{56}$Ni{} mass.
We consider the good match between LL type IIP SN observations and our
results for the light-curve modeling as a strong support for
neutrino-driven explosion models of low-mass, low-energy SN explosions.

\begin{figure*}
\centering
\vspace{5mm}
\includegraphics[width=\textwidth]{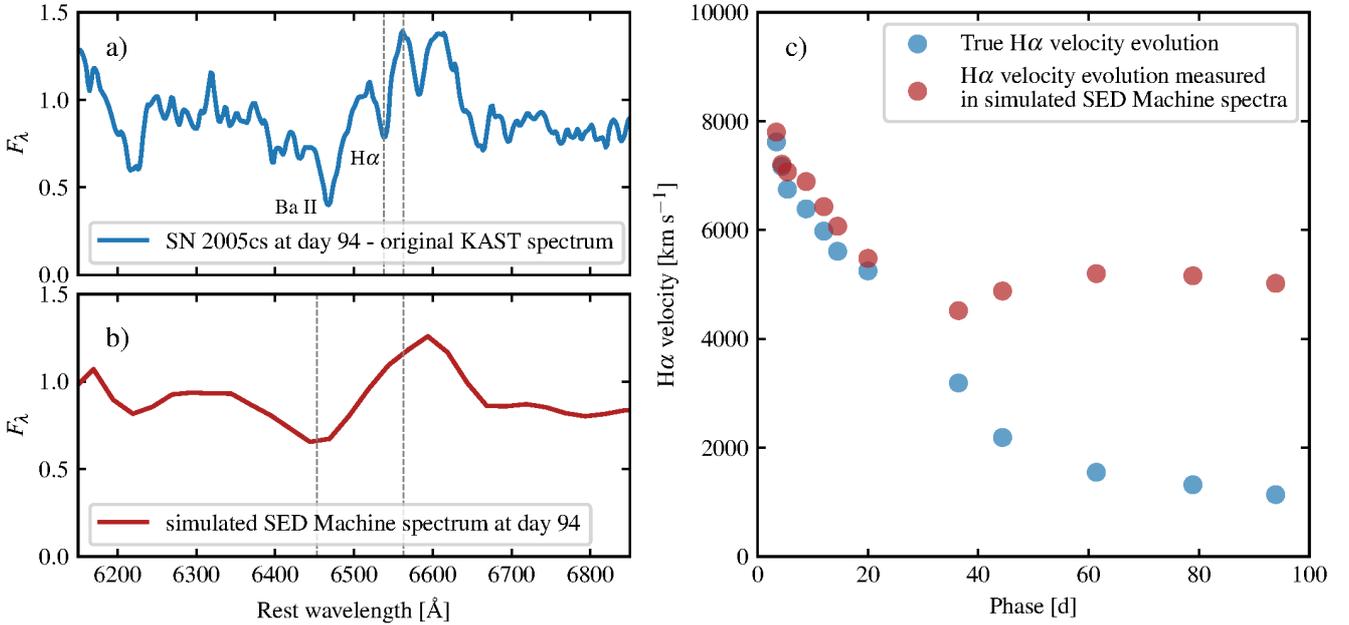}
\caption{Demonstration of the effect of the spectral resolution on the measurement of line velocities. (a) Spectrum of SN~2005cs at 94\,d \citep{2014MNRAS.442..844F}; (b) its artificially degraded analogue 
mimicking the resolution of the P60 spectrograph, and (c) H$\alpha$ velocities deduced from the original spectra compared to those derived from their simulated low-resolution counterparts.}
\label{figure:uph2020cxd}
\end{figure*}

\subsection[Photospheric velocity]{Photospheric velocity}
\label{subsect:uph}

One of the distinct observational features of LL type IIP SNe is the 
relatively low photospheric velocity during the plateau
\citep{2013msao.confE.176P,2014MNRAS.439.2873S}. A low photospheric velocity 
($U_\mathrm{ph}$) or, more precisely, a low ratio between energy and ejecta 
mass \citep{2018MNRAS.475.1937T}, is the strongest diagnostic of low-energy 
explosions.

\subsubsection{Line velocities in SNe 2020cxd and 2005cs}
\label{subsubsect:Uph2020cxd}

In observations, $U_\mathrm{ph}$ cannot be measured directly, but is usually
approximated by measuring the velocity corresponding to the blueshift of the
P-Cygni absorption lines.  \citet{2005AandA...439..671D} have shown, based on
synthetic spectra, that the so-derived velocities are good proxies for
$U_\mathrm{ph}$ in the case of weak lines (such as Fe lines), but that with
a strong line such as H$\alpha$, $U_\mathrm{ph}$ is often overestimated by
up to 40\,\%{}.

For SN~2005cs, \citet{2009MNRAS.394.2266P}  have measured H$\alpha$ and
Sc\,\textsc{ii} velocities that are a factor of $\sim$2 lower than in
normally luminous SNe~II, declining from $\sim$7000\,\kms\ in the earliest
spectra to 1000--1500\,\kms\ at the end of the plateau phase.  The H$\alpha$
velocity evolution determined by \citet{2021AandA...655A..90Y} \blue{and \citet{2022arXiv220303988V}} for SN~2020cxd is
surprisingly different from that of SN~2005cs, starting off at similar
values a few days after the explosion, but declining only to
$\sim$3000\,\kms\ at the end of the plateau.  However, these results have to
be questioned for mainly two reasons:

\begin{enumerate}
\item For epochs later than 90--100\,d, \citet{2021AandA...655A..90Y} \blue{and \citet{2022arXiv220303988V}} chose to measure the full width at half maximum (FWHM)
of the H$\alpha$ emission line rather than the blueshift of the P-Cygni
absorption.  This renders it difficult to directly compare to values determined 
from absorption-line blueshifts.  But even more, there is also a
fundamental problem in the physical interpretation of these numbers.  The
FWHM of an emission line is a good indicator for the size of the emitting
region in fully optically thin conditions, i.e.,  when no part of the
emission is obscured by an optically thick inner core.  However, the
assumption of such conditions would automatically imply that $U_\mathrm{ph}
= 0$, meaning that FWHM measurements of emission lines are never a good way
to measure $U_\mathrm{ph}$.
\item Several SN~2020cxd spectra of \citet{2021AandA...655A..90Y} have been
obtained with the SED Machine spectrograph on the Palomar 60-inch telescope. 
The SED Machine has a very low resolving power of $R =
\frac{\lambda}{\Delta\lambda} \approx 100$ \citep{2018PASP..130c5003B},
corresponding to a velocity resolution of $\Delta v = c \,/\,R \approx
3000$\,\kms.  This sets a lower limit for reliably measurable line
blueshifts that is significantly higher than, e.g., the H$\alpha$ blueshifts
of SN~2005cs towards the end of the plateau.  Consequently, at least the
H$\alpha$ velocity measurement of SN~2020cxd at day 94 ($\sim$3750\,\kms)
appears dubious.  

To assess the possible effect of the SED Machine's low
resolution more quantitatively, we artificially degrade the resolution of a
94\,d spectrum of SN~2005cs \citep{2014MNRAS.442..844F} to yield $R \approx 100$ by
boxcar-smoothing it with a kernel of $\sim$65\,\AA\ and rebinning it to
25\,\AA\ bins. In Figure~\ref{figure:uph2020cxd} we present the result of this
procedure and demonstrate the corresponding effect of the low spectral resolution 
on estimating the photospheric expansion velocities via the H$\alpha$ line as in \citet{2021AandA...655A..90Y}. Panel~(a) shows the spectrum of SN~2005cs taken 
94\,d after the explosion with the KAST spectrograph \citep{2014MNRAS.442..844F}. 
The H$\alpha$ and Ba\,\textsc{ii} lines are rather narrow and well separated. A 
reliable H$\alpha$ velocity can be determined. Panel~(b) displays the same spectrum, 
but artificially degraded to match the resolution of SED Machine ($R\sim100$; \citealt{2018PASP..130c5003B}) at a bin width of 25\,\AA{}. The H$\alpha$ and 
Ba\,\textsc{ii} lines are fully blended, and the velocity inferred from the 
combined absorption trough is several times higher. Panel~(c) presents a 
comparison between the temporal H$\alpha$ velocity evolution of SN~2005cs 
measured in the original spectra \citep{2009MNRAS.394.2266P,2014MNRAS.442..844F} 
and that determined from simulated spectra with SED Machine-like resolution.

In the well-resolved original spectrum, for example, we measure an
H$\alpha$ velocity of $\sim$1150\,\kms, in good agreement with the velocity
evolution shown in \citet{2009MNRAS.394.2266P}.  In the smoothed spectrum,
however, the H$\alpha$ line is completely blended with several
Ba\,\textsc{ii} lines at slightly shorter wavelengths, and an attempt to
measure the blueshift of the resulting trough yields an alleged H$\alpha$
velocity of $\sim$5100\,\kms. If the same Ba\,\textsc{ii} lines also contribute
to the 94\,d spectrum of SN~2020cxd, the reported H$\alpha$ velocity is
likely to be overestimated by a factor of several.
\end{enumerate}

In conclusion, the only line-velocity measurements of SN~2020cxd that can
serve to estimate a photospheric velocity are those at $<$\,40\,d after
explosion, where either the spectra have sufficient resolution or the velocities are high enough that the low resolution of the SED Machine changes the result by at most a few hundred \kms.

\subsubsection{Modelled photospheric velocities}
\label{subsubsect:uphM}

\begin{figure}
\centering
\vspace{5mm}
\includegraphics[width=0.5\textwidth]{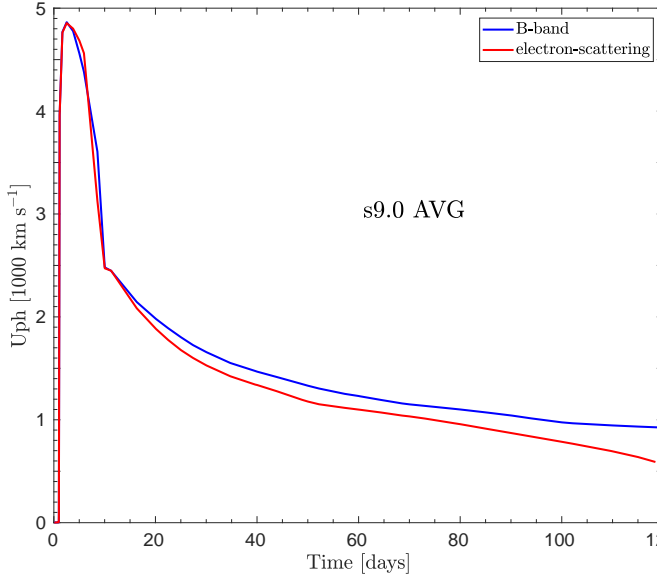}
\caption{Synthetic photospheric velocity estimated via the integrated optical depth in the $B$ band (blue line) compared to that based on the electron-scattering opacity (red line). We use the angle-averaged case of s9.0 for demonstration.}
\label{figure:uphBbandThomson}
\end{figure}

The photospheric velocity in \verb|STELLA| is estimated by default
as the velocity of the shell where the integrated optical depth in the
$B$-band is equal to 2/3. To back up this choice we compare it to the velocity at the location of the electron-scattering photosphere in the case of the angle-averaged explosion model in Figure~\ref{figure:uphBbandThomson}. The integrated Thomson optical depth is computed with the total electron density provided by the Saha equation. In both cases the velocity is calculated at an optical depth equal to 2/3. Figure~\ref{figure:uphBbandThomson} confirms good agreement between the two approaches to estimate the photospheric velocities. The overall behavior is very similar but quantitative differences start at about day 15 and grow slowly with time. The photospheric velocity based on the electron-scattering opacity drops below the B-band value, however it is at most $\sim$20\% smaller until day 100.

\begin{figure} 
\centering
\vspace{5mm}
\includegraphics[width=0.5\textwidth]{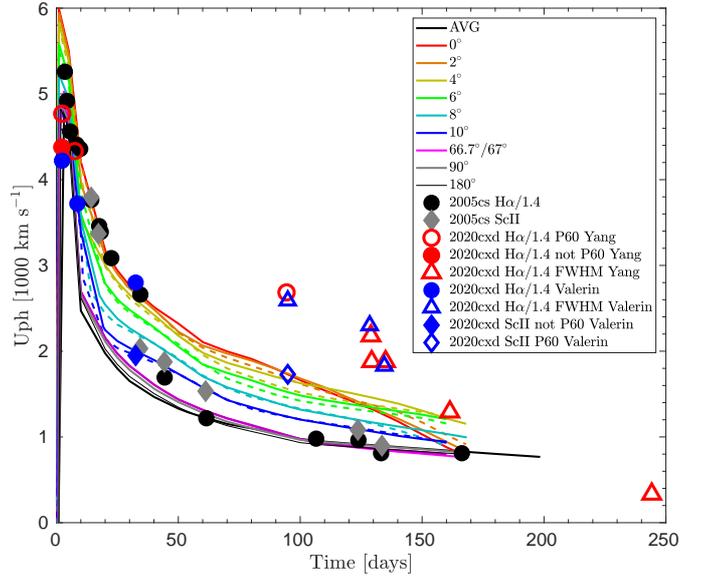}
\caption{ Time evolution of the photospheric velocity, U$_\mathrm{ph}$, for
different radial directions of model s9.0.  The label ``AVG'' corresponds to the angle-averaged profile. The solid and dashed curves indicate the ``$+$'' and ``$-$'' directions for the cases of 2$^\circ$--10$^\circ$, as introduced in Table~\ref{table:rays}. The data points for SN~2005cs are taken from \citet{2009MNRAS.394.2266P}. The red symbols represent data for SN~2020cxd taken from \citet{2021AandA...655A..90Y} and the blue symbols represent data taken from \citet{2022arXiv220303988V}. The H$\alpha$ velocity data are plotted after dividing by a factor of 1.4. The reliable data points are shown as filled symbols, whereas the open symbols represent less reliably measured velocities. See explanation in the text.} 
\label{figure:uph} 
\end{figure}

Figure~\ref{figure:uph} displays the $B$-band photospheric velocity 
evolution for our set of different radial directions from the asymmetric 
3D explosion model s9.0. We superpose the data for the H$\alpha${} and Sc\,\textsc{ii}
velocities of SN~2020cxd and SN~2005cs. We note that the photospheric velocities of SN~2005cs inferred from non-LTE steady-state modelling carried out with CMFGEN \citep[Table~6 in][]{2008ApJ...675..644D} agree well with the observational data points shown in Figure~\ref{figure:uph} (grey and black symbols). The velocity evolution estimated via the
observed H$\alpha${} line in SN~2020cxd \citep{2021AandA...655A..90Y,2022arXiv220303988V} as well as in SN~2005cs
\citep{2009MNRAS.394.2266P} is displayed after dividing the reported
observational data by 1.4. This is considered as 
a proxy to the photospheric velocity, because \citet{2005AandA...439..671D} demonstrated
that the H$\alpha${} velocity exceeds the photospheric velocity by a
factor of up to 1.4. Nevertheless, even after dividing 
by 1.4, the H$\alpha${} velocity
for SN~2020cxd remains too high compared to the values typically found 
for the family of LL type IIP SNe.

The photospheric velocity evolution for all explored radial directions is shown in
Figure~\ref{figure:uph}, because we presume that depending on the viewing direction 
of the SN, intrinsic, large-scale explosion asymmetries of low-mass iron-core progenitors 
as suggested by our explosion model s9.0 from \citet{2020MNRAS.496.2039S} may have some
influence on the observed line velocities, in addition to their possible effects
on the light curve discussed in Section~\ref{subsect:bol}. Therefore, since our 
analysis is based on 1D radiative transfer instead of full 3D transport calculations, 
it may be informative to consider the variation of the photospheric velocities
in the different radial directions.

The spectral observations are sparse and under-resolved during the plateau phase of
SN~2020cxd, as discussed in Section~\ref{subsubsect:Uph2020cxd},
which calls the information available for the photospheric velocity into question. 
Several spectra were taken with the SED Machine on the P60 telescope.
The low spectral resolution of that spectrograph affects ejecta-velocity estimates. 
This is true for the velocities at day~2.4 and day 7.5, and especially at day~94.4,
all of which are probably too high. Moreover, a comparison of the
H$\alpha$ velocities to the photospheric velocity estimated in \verb|STELLA|
has to be taken with caution,
because the synthetic-velocity estimate relies on the
optical depth in the $B$ band and corresponds to the velocity measured via
iron and scandium spectral lines, e.g., Fe\,\textsc{ii}~5169\,\AA{} or
Sc\,\textsc{ii}~6246\,\AA{}. Assuming the H$\alpha$ velocity is correct and
represents the location of the photosphere, the corresponding explosion
energy of SN~2020cxd would be about 1.5~foe\footnote{Assuming the ejecta mass of
9.5~\Msun{} and H$\alpha$ photospheric velocity of 4000~km\,s$^{\,-1}${}, the
diagnostic energy is $M \times v^{\,2}/2 \simeq 1.5\times 10^{\,51}\,\mathrm{erg}=1.5$~foe. However, accounting for the factor of 1.4 in the velocity changes the energy to 0.78~foe.}, 
which is very high even for a canonical type\,IIP SN. Assuming the H$\alpha$ 
velocity is two times higher than the realistic photospheric
velocity (which may be too large a reduction), the explosion energy is
still 0.5~foe, which corresponds to the average explosion energy of core-collapse SNe.

Nevertheless, the spectral lines remain quite narrow later, if we ignore the
resolution constraints of the spectra, 
and at day~240 the measured intrinsic
FWHM of the H$\alpha$ line is only
478~km\,s$^{\,-1}$. Nebular spectral synthesis for the
1D version of model s9.0 displays narrow lines with FWHM about
1000~km\,s$^{\,-1}$ at a similar epoch
\citep{2018MNRAS.475..277J}, corresponding to an explosion energy of 0.11~foe of the
underlying 1D explosion model of s9.0 from \citet{2016ApJ...818..124E} and
\citet{2016ApJ...821...38S}.
The very low velocity of SN 2020cxd at day~240 might be explained by an even lower
explosion energy.
Note that the 1D SN model of s9.0, which was exploded by a parametric
treatment of the neutrino-heating mechanism
\citep{2016ApJ...818..124E,2016ApJ...821...38S}, releases 0.11~foe, whereas the
angle-averaged model of the self-consistent 3D neutrino-driven explosion simulation
\citep{2020MNRAS.496.2039S} has 0.07~foe. This corresponds to a factor of
0.8 in velocity, which is consistent with the observationally diagnosed ejecta 
velocity of SN~2020cxd after about 200~days to some extent.

None of the model results for the different angular directions of the 3D explosion of s9.0 can
explain the photospheric velocity evolution of SN~2020cxd by itself. 
In fact, the H$\alpha${} velocities of SN 2020cxd around 100--150 days exceed the
photospheric velocities of all model directions by far. 
Taking into account the discussion above  and in the previous
Section~\ref{subsubsect:Uph2020cxd}, we compare the synthetic velocities also to the more
reliable data for the LL type IIP SN~2005cs. The photospheric velocity evolution
of SN~2005cs is estimated via the Sc\,\textsc{ii} line \citep{2009MNRAS.394.2266P}.
In contrast to SN~2020cxd, the line velocities of SN~2005cs are close to our computed results or overlap
with them. During the plateau we can explain the photospheric velocity evolution of
SN~2005cs with the velocities for the 8$^\circ${} and 10$^\circ${} directions, while at the
end of the plateau and later our model directions for 66.7$^\circ${},
67$^\circ${}, 90$^\circ${}, 180$^\circ${}, as well as the angle-averaged case match
the observed values. However, we note that the photospheric-velocity estimates
beyond the plateau phase are not perfectly reliable in the simulations carried out
with \verb|STELLA|, because \verb|STELLA| is not capable to provide  
photospheric information when the SN ejecta become semi-transparent.

The fact that the evolution of the observationally diagnosed photospheric velocity of
SN~2005cs agrees with the photosphere of our explosion model in different radial
directions at different times during the plateau on the one hand and with the 
angle-averaged model at the end of the plateau on the other hand, might 
indicate some influence of explosion asymmetries
on the line formation.  This interpretation is compatible with our arguments 
for a possible impact of explosion asymmetries on the light curve of SN~2005cs discussed 
in Section~\ref{subsect:bol}.
Such explosion asymmetries, which are suggested by the low-mass iron-core explosion model s9.0 of
\citet{2020MNRAS.496.2039S}, might play a role also in SN~2020cxd and other low-mass SN explosions.
Unfortunately, no spectrapolarimetric observations have been carried out for SN~2020cxd. 
\citet{2009MNRAS.394.2266P} reported that SN~2005cs possesses asymmetric
H$\alpha${} line shapes during the nebular phase with the emission maximum in the P-Cygni profiles shifted towards redder
wavelengths by 700--800~km\,s$^{\,-1}${}.  The latter might be explained by
an asymmetric distribution of the radioactive
material, i.e., by a dominant fraction of the $^{56}$Ni in the inner part of the SN ejecta pointing
away from the observer \citep{2006AstL...32..739C}. Moreover,
\citet{2007AstL...33..736G} presented imaging polarimetry of
SN~2005cs and reported up to 8\,\% polarisation during the plateau phase.
Such a high degree of polarisation would be unprecedented for Type\,II SNe \citep{2002AJ....124.2506L} and point at an asymmetry in the inner ejecta. However, the inspection of unpublished spectropolarimetric observations of SN~2005cs
obtained with Keck + LRIS (Prog-ID C21L; PI: Leonard) does not seem to
support the results by Gnedin et al.
Another example of a LL type IIP SN with available
spectrapolarimetric observations is SN~2008bk. The plateau
luminosity of this SN is 0.4~dex higher than that of many LL type IIP SNe,
whereas its photospheric velocities are comparable to those of other cases of the low-luminosity family 
\citep{2013msao.confE.176P,2018MNRAS.473.3863L}. 
The polarimetric behaviour during the plateau of SN~2008bk is pretty normal and inconspicuous,
but after the drop from the plateau this SN retains a constant 0.3\,\%{} polarisation, which is unusual
and might indicate some asymmetry inside the ejecta \citep{2012AIPC.1429..204L}.
Therefore, although it is not possible to draw solid conclusions about explosion
asymmetries on grounds of the existing data sets for SN~2005cs and SN~2020cxd,
a possible role of such asymmetries in low-mass, low-energy,
LL type IIP SNe can not be firmly excluded either.  This stresses the need
of more observational data to be compared with predictions for light curves,
spectra, and polarisation based on 3D radiative transfer calculations in 3D
explosion models.

\subsection{Testing the SN progenitor model by Yang et al. 2021}\label{subsect:progenitor}

In order to test the SN properties deduced by \citet{2021AandA...655A..90Y}, we construct two models which we map into
\verb|STELLA|.

The first case is based on model m12 from \citet{2019MNRAS.483.1211K}, which has almost the proposed ejecta mass. Namely, m12 has 11.25~\Msun{} at the pre-collapse stage. Accounting for a mass of 1.45~\Msun{} for the newly-formed compact object results in 9.8~\Msun{} for the SN ejecta. To match the proposed value of 9.5~\Msun{}, we scale the density profile with the factor of 9.5/9.8. The radius of the progenitor is 496~\Rsun{}. To transform this radius to the required value of 187~\Rsun{}, we squeeze each mass zone in the mapped model with a compression factor of 187/496, conserving the mass contained in the zone. We note that model m12 was exploded with the \verb|V1D| code \citep{1993ApJ...412..634L} with an original explosion energy of 0.9~foe. We scale the velocity profile to change this explosion energy to a value of 0.58~foe. The total mass of $^{56}$Ni is scaled to the value of 0.003~\Msun{} for SN~2020cxd. The difference in the $^{56}$Ni mass fraction is added to the silicon abundance.

The second model is based on model s12 from \citet{2016ApJ...821...38S}, which was evolved with the stellar evolution code \verb|KEPLER| until the pre-collapse state. Later, after the necessary modifications, we map the model into \verb|STELLA| and blow it up with the thermal-bomb method, in which a prescribed amount of thermal energy is injected at the inner boundary (i.e., the mass coordinate corresponding to the assumed mass of the compact remnant of 1.35~\Msun{}) within a mass shell of 0.06~\Msun{}.
The total pre-collapse mass of this progenitor is 10.95~\Msun{}. After subtracting 1.35~\Msun{} for the compact object, the ejecta mass is 9.55~\Msun{}, which is very close to the 9.5~\Msun{} value of the analysis by \citet{2021AandA...655A..90Y}. The radius of s12 prior to collapse is 613~\Rsun{}. To make the model more compact, we follow the same procedure as described for model m12 above. The injected thermal-bomb energy is set to 1.73~foe, which leads to a terminal kinetic energy of 0.58~foe. 
We note that these are two different models, computed with different stellar evolution codes. Nevertheless, their chemical structures are very similar, particularly in the hydrogen-rich envelope, which reflects the typical structure of red supergiants and is much more extended than the radius inferred by \citet{2021AandA...655A..90Y}.

\begin{figure}
\centering
\vspace{5mm}
\includegraphics[width=0.5\textwidth]{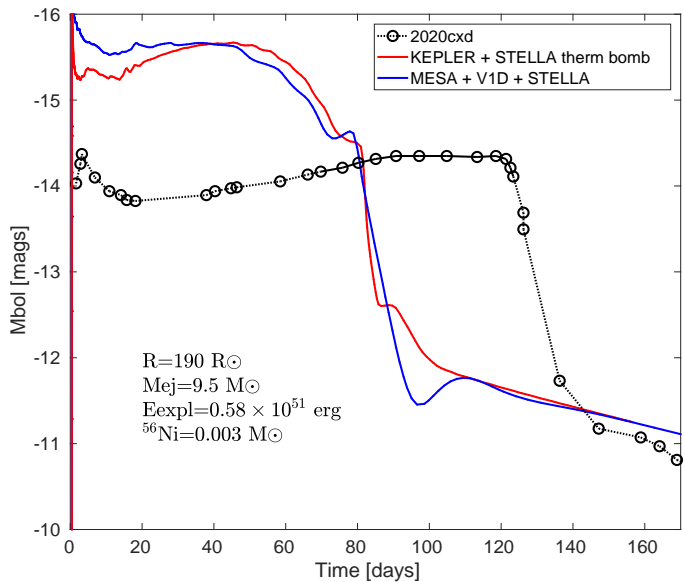}
\caption{Bolometric light curves for our two explosion models s12$_\mathrm{scale}$ and m12$_\mathrm{scale}$ with progenitor and explosion properties adjusted to those inferred by \citet{2021AandA...655A..90Y}.}
\label{figure:m12s12}
\end{figure}

The resulting light curves for the re-scaled m12$_\mathrm{scale}$ and s12$_\mathrm{scale}$ models are shown in Figure~\ref{figure:m12s12}. It is evident that despite m12 and s12 being two different stellar models, exploding them with the same explosion energy after scaling their characteristic progenitor parameters to the same values leads to almost the same bolometric light curves. The explosion energy of 0.58~foe has a distinctive effect on the bolometric luminosity and duration of the plateau. The difference in the light curves at early times is connected to slightly different density structures in the outermost layers of the stellar envelopes. The difference in the light curves at the beginning of the radioactive tail is connected to the numerical viscosity and the zoning in the innermost parts of the ejecta. The plateau of both models lasts only about 80~days versus the 130-day plateau of SN~2020cxd, and has a bolometric magnitude {of }$\,-15.5$~mag, which is 1.5~mag higher than that of SN~2020cxd.

The photospheric velocities are around 2780~km\,s$^{\,-1}$ for both models in the middle of their bolometric plateaus, i.e. around day~40. We note that this value is close to the 2747~km\,s$^{\,-1}$ reported by \citet{2021AandA...655A..90Y} as the expansion velocity of their best-fit model found by the MCMC light curve fitting procedure. This agreement is a consequence of the fact that we set the explosion energy of our models equal to the value inferred by those authors.

With this experiment we show that LL Type IIP SNe, particularly SN~2020cxd, cannot be explained with the progenitor and SN parameters derived by the analysis of \citet{2021AandA...655A..90Y}.

\subsection{Is SN~2020cxd an ECSN?}\label{subsect:ecsn}

\begin{figure*}
\centering
\vspace{5mm}
\includegraphics[width=0.5\textwidth]{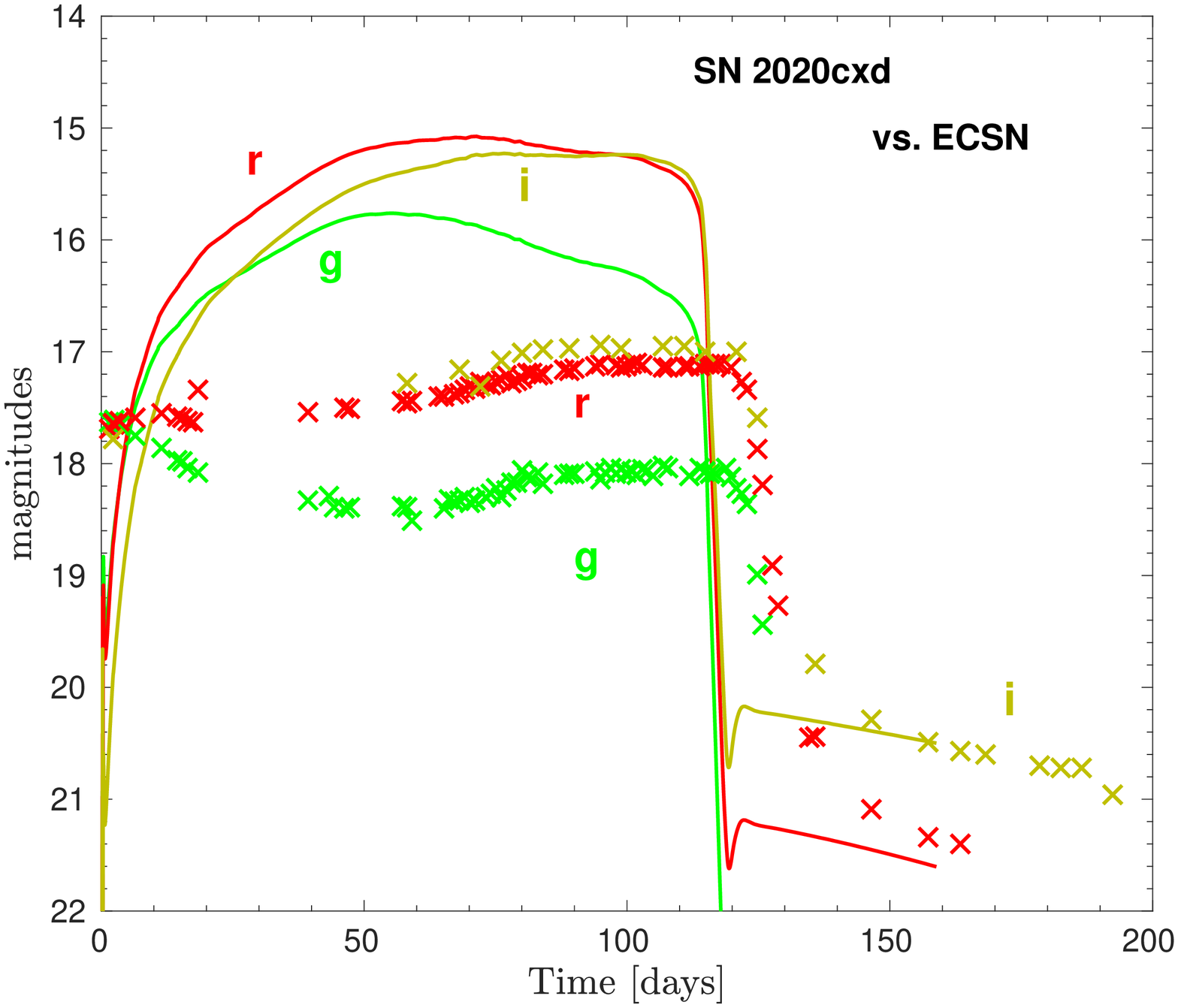}~
\includegraphics[width=0.495\textwidth]{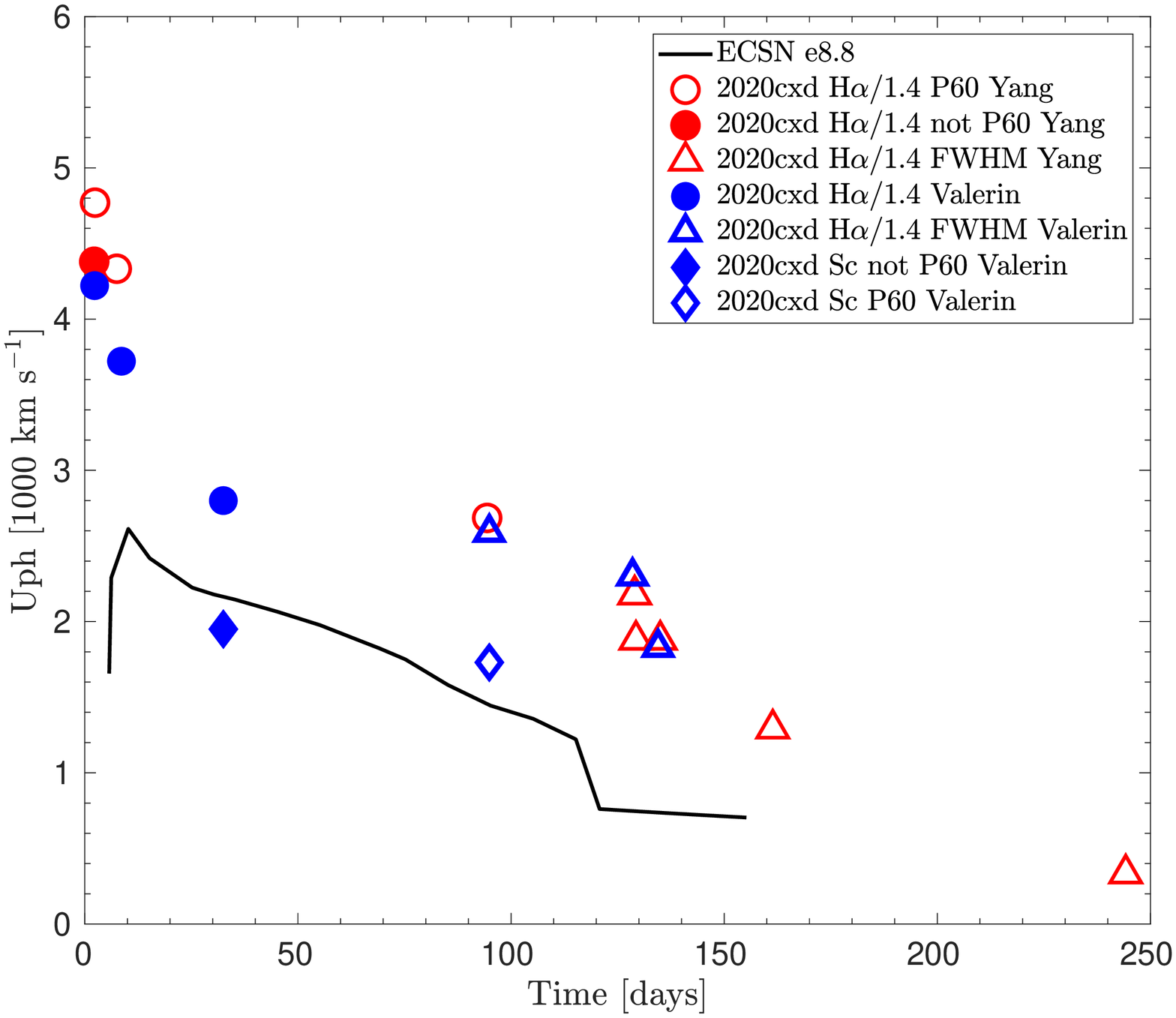}
\caption{Left: $gri$ broad-band light curves for the ECSN model from \citet{2021MNRAS.503..797K} and SN~2020cxd. Right: Photospheric velocity evolution for the ECSN model and SN~2020cxd. The observational data are the same as in Figure~\ref{figure:bands2020cxd} and Figure~\ref{figure:uph}.}
\label{figure:ecsn2020cxd}
\end{figure*}

According to a number of studies, LL type IIP SNe might be interpreted as electron-capture SNe \citep[ECSNe, ][]{2009MNRAS.398.1041B,2009ApJ...705L.138P,2021A&A...654A.157C,2021MNRAS.501.1059R,2022arXiv220303988V}. One of the reasons for this interpretation are identifications of progenitors as low-mass stars with a ZAMS mass around 8~\Msun{}, embedded in some cases in a dusty environment. However, we note that progenitor mass estimates are strongly model dependent with a number of caveats. 

In Figure~\ref{figure:ecsn2020cxd} we show the broad-band magnitudes for SN~2020cxd and for an ECSN model (e8.8) published in \citet{2021MNRAS.503..797K} to test this hypothesis. The progenitor of the ECSN considered in that study has an extended and very tenuous hydrogen-rich envelope. The density and chemical structure of the progenitor are believed to be representative of this kind of massive stars, namely, massive asymptotic giant branch stars. The extended envelope is the result of a sequence of flashes happening during the helium-core burning \citep{2013ApJ...772..150J,2014ApJ...797...83J,2016A&A...593A..72J}. The resulting light curves are far too luminous to be LL type IIP cases, and \citet{2021MNRAS.503..797K} concluded that the characteristic luminosities of ECSNe are comparable to those of reference cases of SNe~IIP. Moreover, ECSNe have a very distinct colour evolution. In particular, the $U$-band magnitude has a plateau at {}$-15.5$~mag during the first 60~days, which is very different from the behaviour of the $U$-band light curve of typical type IIP SNe. The unique $U$-band light curves of ECSNe are explained by the fact that recombination in their envelopes sets in relatively late around day~70. We also superpose the photospheric velocity evolution for our ECSN model e8.8 and the velocities of SN~2020cxd (right plot in Figure~\ref{figure:ecsn2020cxd}). The observed H$\alpha${} velocities (after dividing by 1.4) are roughly 1000~km\,s$^{\,-1}${} higher than those predicted by the ECSN model at an early epoch when the SN~2020cxd velocity estimates are still reliable, although the Sc\,II line velocity estimated at day~32 is consistent with the ECSN.


\section[Conclusions]{Summary and conclusions}
\label{sect:conclusions}


In this study we compared the observational data of the LL type IIP
SNe~2020cxd and 2005cs with multi-band light curves and
photospheric velocities obtained from radiation-hydrodynamics calculations
for a 3D explosion model of a 9.0\,M$_\odot${} (ZAMS mass) red supergiant 
progenitor with an iron core of 1.3\,M$_\odot$ and a radius of 
408~\Rsun{} \citep{2015ApJ...810...34W}. The stellar model had 
a pre-collapse mass of 8.75\,M$_\odot$ (an ejecta mass of 7.4~\Msun{}) and was self-consistently exploded
by the neutrino-driven mechanism in a 3D simulation by \citet{2020MNRAS.496.2039S}.
The model had developed a considerable asphericity by hydrodynamic
instabilities aiding the onset of the explosion. These initial asymmetries
ultimately evolved into large-scale asymmetries in the angular distribution
of {}$^{56}$Ni{}, connected to extended radial mixing of metals 
from the core into the hydrogen-helium envelope. This mixing proceeded
in the form of elongated wide-angle plumes, which led to a prolate
global deformation of the chemical composition in the ejecta
(see Figure~\ref{figure:asymmetry} and for details, see \citealt{2020MNRAS.496.2039S}).

The subsequent long-time SN evolution, starting from the 3D
explosion model at 1.974~days after core bounce, was carried out by
spherically symmetric hydrodynamical simulations with the \verb|STELLA| code,
including multi-band radiative transfer. For that we
considered the spherically averaged 3D model as well as the conditions
in selected radial directions with a different extent of outward metal 
mixing, covering big {}$^{56}$Ni{} plumes as well as regions of less efficient
mixing. Performing radiative transfer calculations with the sphericized 
stellar profiles for these selected directions was intended to demonstrate 
the possible relevance of asymmetry effects in the radiation emission 
of the 3D explosion model. Aspherical radiation transport can, of course,
be reliably treated only by 3D radiative transfer calculations in the
3D SN ejecta, whereas our approach tends to massively overestimate
the influence of direction-dependent variations in the density structure 
and chemical composition. But nevertheless, our calculations might 
demonstrate the basic trends that could be associated with the existence
of such large-scale chemical and density anisotropies in a fully 
multi-dimensional radiative transfer treatment.

We found that our neutrino-driven explosion model of the 9\,M$_\odot${} 
progenitor with an explosion energy of 0.07\,foe and an ejecta
mass of 7.4\,M$_\odot${} can amazingly well reproduce the basic properties
of the bolometric light curve of SN~2020cxd, i.e., its initial decline
to the plateau, the height and duration of the plateau, and the shallow
increase of the plateau luminosity until the steep decline to the 
radioactive-decay tail. This is achieved without any fine tuning of
the explosion model except for a proper scaling of the {}$^{56}$Ni{} mass to
the observationally inferred value of 0.003\,M$_\odot${} for SN~2020cxd.
Such a scaling is motivated by considerable modeling uncertainties with 
respect to an exact determination of the {}$^{56}$Ni{} yield. Good
overall agreement was also obtained for the broad-band light curves
of this SN. In contrast, however, the line velocities reported by
\citet{2021AandA...655A..90Y} do not mirror the time evolution of the photospheric
expansion velocities deduced from our radiative transfer calculations
but are considerably higher (up to a factor of $\sim$3) than the model
values. This holds true even in the directions of the fastest ejecta
expansion connected to the most extended nickel-rich plume, and even 
after scaling down the measured H$\alpha$ line velocities by a factor
of 1.4 as recommended by \citet{2005AandA...439..671D} for comparison with 
the photospheric velocities from radiative transfer modeling.

This is in stark contrast to our findings for SN~2005cs, which is
a template case of a low-energy, LL type IIP SN with low {}$^{56}$Ni{}
production. Here not only the bolometric and broad-band light curves
match the observational ones reasonably well, again without any other
tuning than scaling the {}$^{56}$Ni{} mass. Also the photospheric
velocities are close to the velocities of the Sc\,\textsc{ii} line and the
down-scaled H$\alpha$ line of SN~2005cs and follow their evolutionary 
behavior.

We speculated whether the different shapes of the light
curves of SN~2005cs and SN~2020cxd during the initial decline to the plateau
and the early plateau phase might be connected to large-scale ejecta
asymmetries already at shock breakout and during the early expansion of the 
outer parts of the hydrogen envelope. Such a possibility is suggested by
our 3D explosion model of the 9\,M$_\odot$ progenitor, which exhibits 
large-scale asymmetries of the nickel-rich ejecta, which extend through
the entire hydrogen envelope and lead to asymmetric shock breakout
(see \citealt{2020MNRAS.496.2039S} and Figure~\ref{figure:asymmetry}). 
Contributions by radiation emitted from the faster and more rapidly 
expanding directions of the ejecta might enhance the luminosity of SN~2005cs 
before and during the plateau until about 80--90 days, whereas in SN~2020cxd
such effects might not play an important role or they might not be visible
from our viewing direction. Interestingly, we also
witnessed the trend that at early phases the line velocities of
SN~2005cs were best compatible with the photospheric velocities
in the directions of the fastest ejecta in our aspherical 3D 
explosion model, whereas with progressing time they approached those 
of the ejecta in directions with slower expansion, and, at late times,
they agreed with those of the angle-averaged ejecta. This might
also be interpreted
as a possible indication of large-scale or even global explosion 
asymmetries in the outer ejecta of SN~2005cs, adding to similar conclusions 
previously drawn by \citet{2009MNRAS.394.2266P}, \citet{2006AstL...32..739C}, and
\citet{2007AstL...33..736G} on grounds of asymmetric H$\alpha$ line shapes and 
polarisation measurements. A better theoretical understanding of 
whether such a speculative possibility can explain the light-curve differences 
of SN~2005cs and SN~2020cxd and the evolution of their line velocities will 
require 3D radiative transfer calculations for our 3D explosion model or other 
simulations of asymmetric SNe of low-mass progenitors.

In both of the cases of SN~2020cxd and SN~2005cs, a very low-energy explosion
of our low-mass iron-core progenitor can explain the main light-curve
features. We reason that the similarities between the photospheric velocities
of our model and the line velocities for SN~2005cs and their mismatch in
the case of SN~2020cxd point to a considerable overestimation of the
observationally inferred line velocities in the latter case. We discussed
the corresponding instrumental and diagnostic uncertainties.
Both SNe, therefore, seem to comply with the correlation between explosion
energy and ejected mass of {}$^{56}$Ni{} inferred from Type IIP SN 
observations \citep{2015ApJ...806..225P,2017ApJ...841..127M,2020rfma.book..189P}
and theoretically expected for neutrino-driven explosions 
\citep{2016ApJ...818..124E,2016MNRAS.460..742M,2016ApJ...821...38S,2020ApJ...890...51E}.

A striking difference is observed when comparing the explosion energy of our
explosion and light-curve model, 0.07~foe, to the explosion energies of 
LL type IIP SNe derived with the use of
the so-called Markov Chain Monte Carlo (MCMC) fitting method \citep{2016A&A...589A..53N}.
For example, SN~2020cxd was reported to have a total energy of 0.58~foe \citep{2021AandA...655A..90Y}, SN~2005cs to have a sum of kinetic and
thermal energy of 0.73~foe, and its declared twin PSN~J17292918+7542390
\citep[SN-NGC 6412,][]{2020MNRAS.496.3725J} to have an even higher explosion 
energy of 0.82--0.93~foe. Using the same method,
the reference Type IIP SN~1999em was diagnosed to originate from the explosion 
of a progenitor of 
about 20~\Msun{} (the ejecta mass was estimated to be 19~\Msun{}) with a
diagnostic energy of 4.53~foe. 
At the same time the recent progress in self-consistent
core-collapse explosion simulations coupled with direct 
radiative transfer hydrodynamics calculations for the light curve
has been able to reproduce this SN~1999em by an
explosion of a 15~\Msun{} progenitor with an energy of 0.55~foe \citep{2017ApJ...846...37U}. 
Hence, there is a factor of 10 difference in the derived SN parameters.
The MCMC fitting procedure is likely to suffer from various weak points, 
among them might be a relatively high lower limit for the explosion energy 
of 0.6~foe \citep{2003ApJ...582..905H} and a simplified method for 
computing the bolometric light curve \citep{1989ApJ...340..396A,1993ApJ...414..712P,2014A&A...571A..77N}.
However, the details of the statistical analysis are not fully described
in the mentioned references, which complicates any assessment of the 
employed fitting procedure and results in detail.

\section*{Acknowledgments}
We thank Stephen Justham, Alexei Mironov, Sergei Blinnikov, Patrick Neunteufel, Ferdinando Patat, Stephane Blondin, Luke Shingles, and Ryan Wollager for fruitful discussions.
AK is supported by the Alexander von Humboldt Foundation. 
HTJ and DK acknowledge support by the Deutsche Forschungsgemeinschaft (DFG, German
Research Foundation) through Sonderforschungsbereich (Collaborative Research
Center) SFB-1258 ``Neutrinos and Dark Matter in Astro- and Particle Physics (NDM)'' and under Germany's
Excellence Strategy through Cluster of Excellence ORIGINS (EXC-2094)-390783311, and by the European Research
Council through Grant ERC-AdG No.~341157-COCO2CASA. 
PB is supported  by the grant RSF 21-11-00362.
This research has made use of the Keck Observatory Archive (KOA), which is operated by the W. M. Keck Observatory and the NASA Exoplanet Science Institute (NExScI), under contract with the National Aeronautics and Space Administration.

\addcontentsline{toc}{section}{Acknowledgments}

\section*{Data availability}

The light-curve data computed and analysed for the current study are available via the link
\href{https://wwwmpa.mpa-garching.mpg.de/ccsnarchive/data/Kozyreva2022/}
{https://wwwmpa.mpa-garching.mpg.de/ccsnarchive/data/Kozyreva2022/}.
Angle-averaged data of the 3D core-collapse SN simulation by \citet{2020MNRAS.496.2039S} 
and the radial profiles along the selected angular directions used in the present work are accessible for download
at \href{https://wwwmpa.mpa-garching.mpg.de/ccsnarchive/data/Stockinger2020/}
{https://wwwmpa.mpa-garching.mpg.de/ccsnarchive/data/Stockinger2020/}.

\section*{Software}

\verb|PROMETHEUS-VERTEX| \citep{1989nuas.conf..100F,2002A&A...396..361R,2006A&A...447.1049B};
\verb|PROMETHEUS-HOTB| \citep{2003A&A...408..621K,2006A&A...457..963S,2007A&A...467.1227A,2016ApJ...818..124E}
\verb|NUMPY| and \verb|SCIPY| \citep{Jones2001};
\verb|IPYTHON| \citep{2007CSE.....9c..21P};
\verb|MATPLOTLIB| \citep{2007CSE.....9...90H}; \verb|VisIt| \citep{Childs2012}; \verb|STELLA| \citep{1998ApJ...496..454B,2000ApJ...532.1132B,2006A&A...453..229B}; \verb|IRAF| \citep{1986SPIE..627..733T}.

\bibliographystyle{mnras}
\bibliography{references}

\begin{thebibliography}{}
\makeatletter
\relax
\def\mn@urlcharsother{\let\do\@makeother \do\$\do\&\do\#\do\^\do\_\do\%\do\~}
\def\mn@doi{\begingroup\mn@urlcharsother \@ifnextchar [ {\mn@doi@}
  {\mn@doi@[]}}
\def\mn@doi@[#1]#2{\def\@tempa{#1}\ifx\@tempa\@empty \href
  {http://dx.doi.org/#2} {doi:#2}\else \href {http://dx.doi.org/#2} {#1}\fi
  \endgroup}
\def\mn@eprint#1#2{\mn@eprint@#1:#2::\@nil}
\def\mn@eprint@arXiv#1{\href {http://arxiv.org/abs/#1} {{\tt arXiv:#1}}}
\def\mn@eprint@dblp#1{\href {http://dblp.uni-trier.de/rec/bibtex/#1.xml}
  {dblp:#1}}
\def\mn@eprint@#1:#2:#3:#4\@nil{\def\@tempa {#1}\def\@tempb {#2}\def\@tempc
  {#3}\ifx \@tempc \@empty \let \@tempc \@tempb \let \@tempb \@tempa \fi \ifx
  \@tempb \@empty \def\@tempb {arXiv}\fi \@ifundefined
  {mn@eprint@\@tempb}{\@tempb:\@tempc}{\expandafter \expandafter \csname
  mn@eprint@\@tempb\endcsname \expandafter{\@tempc}}}

\bibitem[\protect\citeauthoryear{{Arcones}, {Janka}  \& {Scheck}}{{Arcones}
  et~al.}{2007}]{2007A&A...467.1227A}
{Arcones} A.,  {Janka} H.~T.,   {Scheck} L.,  2007, \mn@doi [\aap]
  {10.1051/0004-6361:20066983}, \href
  {https://ui.adsabs.harvard.edu/abs/2007A&A...467.1227A} {467, 1227}

\bibitem[\protect\citeauthoryear{{Arnett}}{{Arnett}}{1980}]{1980ApJ...237..541A}
{Arnett} W.~D.,  1980, \mn@doi [\apj] {10.1086/157898}, \href
  {https://ui.adsabs.harvard.edu/abs/1980ApJ...237..541A} {237, 541}

\bibitem[\protect\citeauthoryear{{Arnett} \& {Fu}}{{Arnett} \&
  {Fu}}{1989}]{1989ApJ...340..396A}
{Arnett} W.~D.,  {Fu} A.,  1989, \mn@doi [\apj] {10.1086/167402}, \href
  {http://adsabs.harvard.edu/abs/1989ApJ...340..396A} {340, 396}

\bibitem[\protect\citeauthoryear{{Blagorodnova} et~al.,}{{Blagorodnova}
  et~al.}{2018}]{2018PASP..130c5003B}
{Blagorodnova} N.,  et~al., 2018, \mn@doi [\pasp] {10.1088/1538-3873/aaa53f},
  \href {https://ui.adsabs.harvard.edu/abs/2018PASP..130c5003B} {130, 035003}

\bibitem[\protect\citeauthoryear{{Blinnikov}, {Eastman}, {Bartunov},
  {Popolitov}  \& {Woosley}}{{Blinnikov} et~al.}{1998}]{1998ApJ...496..454B}
{Blinnikov} S.~I.,  {Eastman} R.,  {Bartunov} O.~S.,  {Popolitov} V.~A.,
  {Woosley} S.~E.,  1998, \mn@doi [\apj] {10.1086/305375}, \href
  {http://adsabs.harvard.edu/abs/1998ApJ...496..454B} {496, 454}

\bibitem[\protect\citeauthoryear{{Blinnikov}, {Lundqvist}, {Bartunov}, {Nomoto}
   \& {Iwamoto}}{{Blinnikov} et~al.}{2000}]{2000ApJ...532.1132B}
{Blinnikov} S.,  {Lundqvist} P.,  {Bartunov} O.,  {Nomoto} K.,   {Iwamoto} K.,
  2000, \mn@doi [\apj] {10.1086/308588}, \href
  {http://adsabs.harvard.edu/abs/2000ApJ...532.1132B} {532, 1132}

\bibitem[\protect\citeauthoryear{{Blinnikov}, {R{\"o}pke}, {Sorokina},
  {Gieseler}, {Reinecke}, {Travaglio}, {Hillebrandt}  \&
  {Stritzinger}}{{Blinnikov} et~al.}{2006}]{2006A&A...453..229B}
{Blinnikov} S.~I.,  {R{\"o}pke} F.~K.,  {Sorokina} E.~I.,  {Gieseler} M.,
  {Reinecke} M.,  {Travaglio} C.,  {Hillebrandt} W.,   {Stritzinger} M.,  2006,
  \mn@doi [\aap] {10.1051/0004-6361:20054594}, \href
  {http://adsabs.harvard.edu/abs/2006A%26A...453..229B} {453, 229}

\bibitem[\protect\citeauthoryear{{Bollig}, {Yadav}, {Kresse}, {Janka},
  {M{\"u}ller}  \& {Heger}}{{Bollig} et~al.}{2021}]{2021ApJ...915...28B}
{Bollig} R.,  {Yadav} N.,  {Kresse} D.,  {Janka} H.-T.,  {M{\"u}ller} B.,
  {Heger} A.,  2021, \mn@doi [\apj] {10.3847/1538-4357/abf82e}, \href
  {https://ui.adsabs.harvard.edu/abs/2021ApJ...915...28B} {915, 28}

\bibitem[\protect\citeauthoryear{{Botticella} et~al.,}{{Botticella}
  et~al.}{2009}]{2009MNRAS.398.1041B}
{Botticella} M.~T.,  et~al., 2009, \mn@doi [\mnras]
  {10.1111/j.1365-2966.2009.15082.x}, \href
  {https://ui.adsabs.harvard.edu/abs/2009MNRAS.398.1041B} {398, 1041}

\bibitem[\protect\citeauthoryear{{Buras}, {Rampp}, {Janka}  \&
  {Kifonidis}}{{Buras} et~al.}{2006}]{2006A&A...447.1049B}
{Buras} R.,  {Rampp} M.,  {Janka} H.~T.,   {Kifonidis} K.,  2006, \mn@doi
  [\aap] {10.1051/0004-6361:20053783}, \href
  {https://ui.adsabs.harvard.edu/abs/2006A&A...447.1049B} {447, 1049}

\bibitem[\protect\citeauthoryear{{Burrows}, {Radice}  \& {Vartanyan}}{{Burrows}
  et~al.}{2019}]{2019MNRAS.485.3153B}
{Burrows} A.,  {Radice} D.,   {Vartanyan} D.,  2019, \mn@doi [\mnras]
  {10.1093/mnras/stz543}, \href
  {https://ui.adsabs.harvard.edu/abs/2019MNRAS.485.3153B} {485, 3153}

\bibitem[\protect\citeauthoryear{{Cai} et~al.,}{{Cai}
  et~al.}{2021}]{2021A&A...654A.157C}
{Cai} Y.~Z.,  et~al., 2021, \mn@doi [\aap] {10.1051/0004-6361/202141078}, \href
  {https://ui.adsabs.harvard.edu/abs/2021A&A...654A.157C} {654, A157}

\bibitem[\protect\citeauthoryear{{Caputo}, {Janka}, {Raffelt}  \&
  {Vitagliano}}{{Caputo} et~al.}{2022a}]{Caputo+2022}
{Caputo} A.,  {Janka} H.-T.,  {Raffelt} G.,   {Vitagliano} E.,  2022a, arXiv
  e-prints, \href {https://ui.adsabs.harvard.edu/abs/2022arXiv220109890C} {p.
  arXiv:2201.09890}

\bibitem[\protect\citeauthoryear{{Caputo}, {Raffelt}  \& {Vitagliano}}{{Caputo}
  et~al.}{2022b}]{Caputo+2022a}
{Caputo} A.,  {Raffelt} G.,   {Vitagliano} E.,  2022b, \mn@doi [\prd]
  {10.1103/PhysRevD.105.035022}, \href
  {https://ui.adsabs.harvard.edu/abs/2022PhRvD.105c5022C} {105, 035022}

\bibitem[\protect\citeauthoryear{{Carenza}, {Fischer}, {Giannotti}, {Guo},
  {Mart{\'\i}nez-Pinedo}  \& {Mirizzi}}{{Carenza}
  et~al.}{2019}]{2019JCAP...10..016C}
{Carenza} P.,  {Fischer} T.,  {Giannotti} M.,  {Guo} G.,
  {Mart{\'\i}nez-Pinedo} G.,   {Mirizzi} A.,  2019, \mn@doi [\jcap]
  {10.1088/1475-7516/2019/10/016}, \href
  {https://ui.adsabs.harvard.edu/abs/2019JCAP...10..016C} {2019, 016}

\bibitem[\protect\citeauthoryear{{Chang}, {Essig}  \& {McDermott}}{{Chang}
  et~al.}{2018}]{2018JHEP...09..051C}
{Chang} J.~H.,  {Essig} R.,   {McDermott} S.~D.,  2018, \mn@doi [Journal of
  High Energy Physics] {10.1007/JHEP09(2018)051}, \href
  {https://ui.adsabs.harvard.edu/abs/2018JHEP...09..051C} {2018, 51}

\bibitem[\protect\citeauthoryear{{Childs} et~al.,}{{Childs}
  et~al.}{2012}]{Childs2012}
{Childs} H.,  et~al., 2012, in {Wes Bethel} E.,  {Childs} H.,   {Hansen} C.,
  eds, High Performance Visualization--Enabling Extreme-Scale Scientific
  Insight. Boca Raton, FL: CRC Press, pp 357--372

\bibitem[\protect\citeauthoryear{{Chugai}}{{Chugai}}{2006}]{2006AstL...32..739C}
{Chugai} N.~N.,  2006, \mn@doi [Astronomy Letters] {10.1134/S1063773706110041},
  \href {https://ui.adsabs.harvard.edu/abs/2006AstL...32..739C} {32, 739}

\bibitem[\protect\citeauthoryear{{Dessart} \& {Hillier}}{{Dessart} \&
  {Hillier}}{2005}]{2005AandA...439..671D}
{Dessart} L.,  {Hillier} D.~J.,  2005, \mn@doi [\aap]
  {10.1051/0004-6361:20053217}, \href
  {https://ui.adsabs.harvard.edu/abs/2005A&A...439..671D} {439, 671}

\bibitem[\protect\citeauthoryear{{Dessart} et~al.,}{{Dessart}
  et~al.}{2008}]{2008ApJ...675..644D}
{Dessart} L.,  et~al., 2008, \mn@doi [\apj] {10.1086/526451}, \href
  {http://adsabs.harvard.edu/abs/2008ApJ...675..644D} {675, 644}

\bibitem[\protect\citeauthoryear{{Ertl}, {Janka}, {Woosley}, {Sukhbold}  \&
  {Ugliano}}{{Ertl} et~al.}{2016}]{2016ApJ...818..124E}
{Ertl} T.,  {Janka} H.~T.,  {Woosley} S.~E.,  {Sukhbold} T.,   {Ugliano} M.,
  2016, \mn@doi [\apj] {10.3847/0004-637X/818/2/124}, \href
  {https://ui.adsabs.harvard.edu/abs/2016ApJ...818..124E} {818, 124}

\bibitem[\protect\citeauthoryear{{Ertl}, {Woosley}, {Sukhbold}  \&
  {Janka}}{{Ertl} et~al.}{2020}]{2020ApJ...890...51E}
{Ertl} T.,  {Woosley} S.~E.,  {Sukhbold} T.,   {Janka} H.~T.,  2020, \mn@doi
  [\apj] {10.3847/1538-4357/ab6458}, \href
  {https://ui.adsabs.harvard.edu/abs/2020ApJ...890...51E} {890, 51}

\bibitem[\protect\citeauthoryear{{Faran} et~al.,}{{Faran}
  et~al.}{2014}]{2014MNRAS.442..844F}
{Faran} T.,  et~al., 2014, \mn@doi [\mnras] {10.1093/mnras/stu955}, \href
  {http://adsabs.harvard.edu/abs/2014MNRAS.442..844F} {442, 844}

\bibitem[\protect\citeauthoryear{{Fryxell}, {M{\"u}ller}  \&
  {Arnett}}{{Fryxell} et~al.}{1989}]{1989nuas.conf..100F}
{Fryxell} B.,  {M{\"u}ller} E.,   {Arnett} D.,  1989, in {Hillebrandt} W.,
  {M{\"u}ller} E.,  eds, Nuclear Astrophysics.

\bibitem[\protect\citeauthoryear{{Glas}, {Just}, {Janka}  \&
  {Obergaulinger}}{{Glas} et~al.}{2019}]{2019ApJ...873...45G}
{Glas} R.,  {Just} O.,  {Janka} H.~T.,   {Obergaulinger} M.,  2019, \mn@doi
  [\apj] {10.3847/1538-4357/ab0423}, \href
  {https://ui.adsabs.harvard.edu/abs/2019ApJ...873...45G} {873, 45}

\bibitem[\protect\citeauthoryear{{Gnedin}, {Larionov}, {Konstantinova}  \&
  {Kopatskaya}}{{Gnedin} et~al.}{2007}]{2007AstL...33..736G}
{Gnedin} Y.~N.,  {Larionov} V.~M.,  {Konstantinova} T.~S.,   {Kopatskaya}
  E.~N.,  2007, \mn@doi [Astronomy Letters] {10.1134/S1063773707110047}, \href
  {https://ui.adsabs.harvard.edu/abs/2007AstL...33..736G} {33, 736}

\bibitem[\protect\citeauthoryear{{Goldberg}, {Bildsten}  \&
  {Paxton}}{{Goldberg} et~al.}{2019}]{2019ApJ...879....3G}
{Goldberg} J.~A.,  {Bildsten} L.,   {Paxton} B.,  2019, \mn@doi [\apj]
  {10.3847/1538-4357/ab22b6}, \href
  {https://ui.adsabs.harvard.edu/abs/2019ApJ...879....3G} {879, 3}

\bibitem[\protect\citeauthoryear{{Grasberg} \& {Nadezhin}}{{Grasberg} \&
  {Nadezhin}}{1976}]{1976Ap&SS..44..409G}
{Grasberg} E.~K.,  {Nadezhin} D.~K.,  1976, \mn@doi [\apss]
  {10.1007/BF00642529}, \href
  {http://adsabs.harvard.edu/abs/1976Ap%26SS..44..409G} {44, 409}

\bibitem[\protect\citeauthoryear{{Hamuy}}{{Hamuy}}{2003}]{2003ApJ...582..905H}
{Hamuy} M.,  2003, \mn@doi [\apj] {10.1086/344689}, \href
  {http://adsabs.harvard.edu/abs/2003ApJ...582..905H} {582, 905}

\bibitem[\protect\citeauthoryear{{Hunter}}{{Hunter}}{2007}]{2007CSE.....9...90H}
{Hunter} J.~D.,  2007, \mn@doi [Computing in Science and Engineering]
  {10.1109/MCSE.2007.55}, \href
  {https://ui.adsabs.harvard.edu/abs/2007CSE.....9...90H} {9, 90}

\bibitem[\protect\citeauthoryear{{J{\"a}ger} Zolt{\'a}n et~al.,}{{J{\"a}ger}
  et~al.}{2020}]{2020MNRAS.496.3725J}
{J{\"a}ger} Zolt{\'a}n J.,  et~al., 2020, \mn@doi [\mnras]
  {10.1093/mnras/staa1743}, \href
  {https://ui.adsabs.harvard.edu/abs/2020MNRAS.496.3725J} {496, 3725}

\bibitem[\protect\citeauthoryear{{Janka}}{{Janka}}{2017}]{2017hsn..book.1095J}
{Janka} H.-T.,  2017, in {Alsabti} A.~W.,  {Murdin} P.,  eds, , Handbook of
  Supernovae.
Cham, Switzerland: Springer Nature, p.~1095,
  \mn@doi{10.1007/978-3-319-21846-5\_109}

\bibitem[\protect\citeauthoryear{{Jerkstrand}, {Ertl}, {Janka}, {M{\"u}ller},
  {Sukhbold}  \& {Woosley}}{{Jerkstrand} et~al.}{2018}]{2018MNRAS.475..277J}
{Jerkstrand} A.,  {Ertl} T.,  {Janka} H.~T.,  {M{\"u}ller} E.,  {Sukhbold} T.,
   {Woosley} S.~E.,  2018, \mn@doi [\mnras] {10.1093/mnras/stx2877}, \href
  {https://ui.adsabs.harvard.edu/abs/2018MNRAS.475..277J} {475, 277}

\bibitem[\protect\citeauthoryear{Jones, Oliphant, Peterson  et~al.}{Jones
  et~al.}{2001}]{Jones2001}
Jones E.,  Oliphant T.,  Peterson P.,   et~al., 2001, {SciPy}: Open source
  scientific tools for {Python}, \url {http://www.scipy.org/}

\bibitem[\protect\citeauthoryear{{Jones} et~al.,}{{Jones}
  et~al.}{2013}]{2013ApJ...772..150J}
{Jones} S.,  et~al., 2013, \mn@doi [\apj] {10.1088/0004-637X/772/2/150}, \href
  {https://ui.adsabs.harvard.edu/abs/2013ApJ...772..150J} {772, 150}

\bibitem[\protect\citeauthoryear{{Jones}, {Hirschi}  \& {Nomoto}}{{Jones}
  et~al.}{2014}]{2014ApJ...797...83J}
{Jones} S.,  {Hirschi} R.,   {Nomoto} K.,  2014, \mn@doi [\apj]
  {10.1088/0004-637X/797/2/83}, \href
  {https://ui.adsabs.harvard.edu/abs/2014ApJ...797...83J} {797, 83}

\bibitem[\protect\citeauthoryear{{Jones}, {R{\"o}pke}, {Pakmor}, {Seitenzahl},
  {Ohlmann}  \& {Edelmann}}{{Jones} et~al.}{2016}]{2016A&A...593A..72J}
{Jones} S.,  {R{\"o}pke} F.~K.,  {Pakmor} R.,  {Seitenzahl} I.~R.,  {Ohlmann}
  S.~T.,   {Edelmann} P.~V.~F.,  2016, \mn@doi [\aap]
  {10.1051/0004-6361/201628321}, \href
  {https://ui.adsabs.harvard.edu/abs/2016A&A...593A..72J} {593, A72}

\bibitem[\protect\citeauthoryear{{Kasen} \& {Woosley}}{{Kasen} \&
  {Woosley}}{2009}]{2009ApJ...703.2205K}
{Kasen} D.,  {Woosley} S.~E.,  2009, \mn@doi [\apj]
  {10.1088/0004-637X/703/2/2205}, \href
  {http://adsabs.harvard.edu/abs/2009ApJ...703.2205K} {703, 2205}

\bibitem[\protect\citeauthoryear{{Kifonidis}, {Plewa}, {Janka}  \&
  {M{\"u}ller}}{{Kifonidis} et~al.}{2003}]{2003A&A...408..621K}
{Kifonidis} K.,  {Plewa} T.,  {Janka} H.~T.,   {M{\"u}ller} E.,  2003, \mn@doi
  [\aap] {10.1051/0004-6361:20030863}, \href
  {https://ui.adsabs.harvard.edu/abs/2003A&A...408..621K} {408, 621}

\bibitem[\protect\citeauthoryear{{Kozyreva}, {Nakar}  \& {Waldman}}{{Kozyreva}
  et~al.}{2019}]{2019MNRAS.483.1211K}
{Kozyreva} A.,  {Nakar} E.,   {Waldman} R.,  2019, \mn@doi [\mnras]
  {10.1093/mnras/sty3185}, \href
  {http://adsabs.harvard.edu/abs/2019MNRAS.483.1211K} {483, 1211}

\bibitem[\protect\citeauthoryear{{Kozyreva}, {Shingles}, {Mironov}, {Baklanov}
  \& {Blinnikov}}{{Kozyreva} et~al.}{2020}]{2020MNRAS.499.4312K}
{Kozyreva} A.,  {Shingles} L.,  {Mironov} A.,  {Baklanov} P.,   {Blinnikov} S.,
   2020, \mn@doi [\mnras] {10.1093/mnras/staa2704}, \href
  {https://ui.adsabs.harvard.edu/abs/2020MNRAS.499.4312K} {499, 4312}

\bibitem[\protect\citeauthoryear{{Kozyreva}, {Baklanov}, {Jones}, {Stockinger}
  \& {Janka}}{{Kozyreva} et~al.}{2021}]{2021MNRAS.503..797K}
{Kozyreva} A.,  {Baklanov} P.,  {Jones} S.,  {Stockinger} G.,   {Janka} H.-T.,
  2021, \mn@doi [\mnras] {10.1093/mnras/stab350}, \href
  {https://ui.adsabs.harvard.edu/abs/2021MNRAS.503..797K} {503, 797}

\bibitem[\protect\citeauthoryear{{Lentz} et~al.,}{{Lentz}
  et~al.}{2015}]{2015ApJ...807L..31L}
{Lentz} E.~J.,  et~al., 2015, \mn@doi [\apjl] {10.1088/2041-8205/807/2/L31},
  \href {https://ui.adsabs.harvard.edu/abs/2015ApJ...807L..31L} {807, L31}

\bibitem[\protect\citeauthoryear{{Leonard}, {Filippenko}, {Chornock}  \&
  {Li}}{{Leonard} et~al.}{2002}]{2002AJ....124.2506L}
{Leonard} D.~C.,  {Filippenko} A.~V.,  {Chornock} R.,   {Li} W.,  2002, \mn@doi
  [\aj] {10.1086/343772}, \href
  {https://ui.adsabs.harvard.edu/abs/2002AJ....124.2506L} {124, 2506}

\bibitem[\protect\citeauthoryear{{Leonard}, {Dessart}, {Hillier}  \&
  {Pignata}}{{Leonard} et~al.}{2012}]{2012AIPC.1429..204L}
{Leonard} D.~C.,  {Dessart} L.,  {Hillier} D.~J.,   {Pignata} G.,  2012, in
  {Hoffman} J.~L.,  {Bjorkman} J.,   {Whitney} B.,  eds,  American Institute of
  Physics Conference Series Vol. 1429, Stellar Polarimetry: from Birth to
  Death. pp 204--207 (\mn@eprint {arXiv} {1109.5406}),
  \mn@doi{10.1063/1.3701926}

\bibitem[\protect\citeauthoryear{{Levesque}, {Massey}, {Olsen}, {Plez},
  {Josselin}, {Maeder}  \& {Meynet}}{{Levesque}
  et~al.}{2005}]{2005ApJ...628..973L}
{Levesque} E.~M.,  {Massey} P.,  {Olsen} K.~A.~G.,  {Plez} B.,  {Josselin} E.,
  {Maeder} A.,   {Meynet} G.,  2005, \mn@doi [\apj] {10.1086/430901}, \href
  {https://ui.adsabs.harvard.edu/abs/2005ApJ...628..973L} {628, 973}

\bibitem[\protect\citeauthoryear{{Levesque}, {Massey}, {Olsen}, {Plez},
  {Meynet}  \& {Maeder}}{{Levesque} et~al.}{2006}]{2006ApJ...645.1102L}
{Levesque} E.~M.,  {Massey} P.,  {Olsen} K.~A.~G.,  {Plez} B.,  {Meynet} G.,
  {Maeder} A.,  2006, \mn@doi [\apj] {10.1086/504417}, \href
  {https://ui.adsabs.harvard.edu/abs/2006ApJ...645.1102L} {645, 1102}

\bibitem[\protect\citeauthoryear{{Lisakov}, {Dessart}, {Hillier}, {Waldman}  \&
  {Livne}}{{Lisakov} et~al.}{2018}]{2018MNRAS.473.3863L}
{Lisakov} S.~M.,  {Dessart} L.,  {Hillier} D.~J.,  {Waldman} R.,   {Livne} E.,
  2018, \mn@doi [\mnras] {10.1093/mnras/stx2521}, \href
  {https://ui.adsabs.harvard.edu/abs/2018MNRAS.473.3863L} {473, 3863}

\bibitem[\protect\citeauthoryear{{Livne}}{{Livne}}{1993}]{1993ApJ...412..634L}
{Livne} E.,  1993, \mn@doi [\apj] {10.1086/172950}, \href
  {http://adsabs.harvard.edu/abs/1993ApJ...412..634L} {412, 634}

\bibitem[\protect\citeauthoryear{{Mattila}, {Smartt}, {Eldridge}, {Maund},
  {Crockett}  \& {Danziger}}{{Mattila} et~al.}{2008}]{2008ApJ...688L..91M}
{Mattila} S.,  {Smartt} S.~J.,  {Eldridge} J.~J.,  {Maund} J.~R.,  {Crockett}
  R.~M.,   {Danziger} I.~J.,  2008, \mn@doi [\apjl] {10.1086/595587}, \href
  {https://ui.adsabs.harvard.edu/abs/2008ApJ...688L..91M} {688, L91}

\bibitem[\protect\citeauthoryear{{Melson}, {Janka}  \& {Marek}}{{Melson}
  et~al.}{2015a}]{2015ApJ...801L..24M}
{Melson} T.,  {Janka} H.-T.,   {Marek} A.,  2015a, \mn@doi [\apjl]
  {10.1088/2041-8205/801/2/L24}, \href
  {https://ui.adsabs.harvard.edu/abs/2015ApJ...801L..24M} {801, L24}

\bibitem[\protect\citeauthoryear{{Melson}, {Janka}, {Bollig}, {Hanke}, {Marek}
  \& {M{\"u}ller}}{{Melson} et~al.}{2015b}]{2015ApJ...808L..42M}
{Melson} T.,  {Janka} H.-T.,  {Bollig} R.,  {Hanke} F.,  {Marek} A.,
  {M{\"u}ller} B.,  2015b, \mn@doi [\apjl] {10.1088/2041-8205/808/2/L42}, \href
  {https://ui.adsabs.harvard.edu/abs/2015ApJ...808L..42M} {808, L42}

\bibitem[\protect\citeauthoryear{{Melson}, {Kresse}  \& {Janka}}{{Melson}
  et~al.}{2020}]{2020ApJ...891...27M}
{Melson} T.,  {Kresse} D.,   {Janka} H.-T.,  2020, \mn@doi [\apj]
  {10.3847/1538-4357/ab72a7}, \href
  {https://ui.adsabs.harvard.edu/abs/2020ApJ...891...27M} {891, 27}

\bibitem[\protect\citeauthoryear{{Mori}, {Takiwaki}, {Kotake}  \&
  {Horiuchi}}{{Mori} et~al.}{2022}]{Mori+2022}
{Mori} K.,  {Takiwaki} T.,  {Kotake} K.,   {Horiuchi} S.,  2022, \mn@doi
  [Physical Review D] {10.1103/PhysRevD.105.063009}, \href
  {https://ui.adsabs.harvard.edu/abs/2022PhRvD.105f3009M} {105, 063009}

\bibitem[\protect\citeauthoryear{{Moriya}, {Suzuki}, {Takiwaki}, {Pan}  \&
  {Blinnikov}}{{Moriya} et~al.}{2020}]{2020MNRAS.497.1619M}
{Moriya} T.~J.,  {Suzuki} A.,  {Takiwaki} T.,  {Pan} Y.-C.,   {Blinnikov}
  S.~I.,  2020, \mn@doi [\mnras] {10.1093/mnras/staa2060}, \href
  {https://ui.adsabs.harvard.edu/abs/2020MNRAS.497.1619M} {497, 1619}

\bibitem[\protect\citeauthoryear{{M{\"u}ller}, {Heger}, {Liptai}  \&
  {Cameron}}{{M{\"u}ller} et~al.}{2016}]{2016MNRAS.460..742M}
{M{\"u}ller} B.,  {Heger} A.,  {Liptai} D.,   {Cameron} J.~B.,  2016, \mn@doi
  [\mnras] {10.1093/mnras/stw1083}, \href
  {https://ui.adsabs.harvard.edu/abs/2016MNRAS.460..742M} {460, 742}

\bibitem[\protect\citeauthoryear{{M{\"u}ller}, {Melson}, {Heger}  \&
  {Janka}}{{M{\"u}ller} et~al.}{2017a}]{2017MNRAS.472..491M}
{M{\"u}ller} B.,  {Melson} T.,  {Heger} A.,   {Janka} H.-T.,  2017a, \mn@doi
  [\mnras] {10.1093/mnras/stx1962}, \href
  {https://ui.adsabs.harvard.edu/abs/2017MNRAS.472..491M} {472, 491}

\bibitem[\protect\citeauthoryear{{M{\"u}ller}, {Prieto}, {Pejcha}  \&
  {Clocchiatti}}{{M{\"u}ller} et~al.}{2017b}]{2017ApJ...841..127M}
{M{\"u}ller} T.,  {Prieto} J.~L.,  {Pejcha} O.,   {Clocchiatti} A.,  2017b,
  \mn@doi [\apj] {10.3847/1538-4357/aa72f1}, \href
  {http://adsabs.harvard.edu/abs/2017ApJ...841..127M} {841, 127}

\bibitem[\protect\citeauthoryear{{M{\"u}ller}, {Gay}, {Heger}, {Tauris}  \&
  {Sim}}{{M{\"u}ller} et~al.}{2018}]{2018MNRAS.479.3675M}
{M{\"u}ller} B.,  {Gay} D.~W.,  {Heger} A.,  {Tauris} T.~M.,   {Sim} S.~A.,
  2018, \mn@doi [\mnras] {10.1093/mnras/sty1683}, \href
  {https://ui.adsabs.harvard.edu/abs/2018MNRAS.479.3675M} {479, 3675}

\bibitem[\protect\citeauthoryear{{Nagy} \& {Vink{\'o}}}{{Nagy} \&
  {Vink{\'o}}}{2016}]{2016A&A...589A..53N}
{Nagy} A.~P.,  {Vink{\'o}} J.,  2016, \mn@doi [\aap]
  {10.1051/0004-6361/201527931}, \href
  {https://ui.adsabs.harvard.edu/abs/2016A&A...589A..53N} {589, A53}

\bibitem[\protect\citeauthoryear{{Nagy}, {Ordasi}, {Vink{\'o}}  \&
  {Wheeler}}{{Nagy} et~al.}{2014}]{2014A&A...571A..77N}
{Nagy} A.~P.,  {Ordasi} A.,  {Vink{\'o}} J.,   {Wheeler} J.~C.,  2014, \mn@doi
  [\aap] {10.1051/0004-6361/201424237}, \href
  {https://ui.adsabs.harvard.edu/abs/2014A&A...571A..77N} {571, A77}

\bibitem[\protect\citeauthoryear{{Nakamura}, {Takiwaki}, {Kuroda}  \&
  {Kotake}}{{Nakamura} et~al.}{2015}]{2015PASJ...67..107N}
{Nakamura} K.,  {Takiwaki} T.,  {Kuroda} T.,   {Kotake} K.,  2015, \mn@doi
  [\pasj] {10.1093/pasj/psv073}, \href
  {https://ui.adsabs.harvard.edu/abs/2015PASJ...67..107N} {67, 107}

\bibitem[\protect\citeauthoryear{{O'Connor} \& {Ott}}{{O'Connor} \&
  {Ott}}{2011}]{2011ApJ...730...70O}
{O'Connor} E.,  {Ott} C.~D.,  2011, \mn@doi [\apj]
  {10.1088/0004-637X/730/2/70}, \href
  {https://ui.adsabs.harvard.edu/abs/2011ApJ...730...70O} {730, 70}

\bibitem[\protect\citeauthoryear{{Ott}, {Roberts}, {da Silva Schneider},
  {Fedrow}, {Haas}  \& {Schnetter}}{{Ott} et~al.}{2018}]{2018ApJ...855L...3O}
{Ott} C.~D.,  {Roberts} L.~F.,  {da Silva Schneider} A.,  {Fedrow} J.~M.,
  {Haas} R.,   {Schnetter} E.,  2018, \mn@doi [\apjl]
  {10.3847/2041-8213/aaa967}, \href
  {https://ui.adsabs.harvard.edu/abs/2018ApJ...855L...3O} {855, L3}

\bibitem[\protect\citeauthoryear{{Pastorello} et~al.,}{{Pastorello}
  et~al.}{2006}]{2006MNRAS.370.1752P}
{Pastorello} A.,  et~al., 2006, \mn@doi [\mnras]
  {10.1111/j.1365-2966.2006.10587.x}, \href
  {https://ui.adsabs.harvard.edu/abs/2006MNRAS.370.1752P} {370, 1752}

\bibitem[\protect\citeauthoryear{{Pastorello} et~al.,}{{Pastorello}
  et~al.}{2009}]{2009MNRAS.394.2266P}
{Pastorello} A.,  et~al., 2009, \mn@doi [\mnras]
  {10.1111/j.1365-2966.2009.14505.x}, \href
  {https://ui.adsabs.harvard.edu/abs/2009MNRAS.394.2266P} {394, 2266}

\bibitem[\protect\citeauthoryear{{Pejcha}}{{Pejcha}}{2020}]{2020rfma.book..189P}
{Pejcha} O.,  2020, in {Kab{\'a}th} P.,  {Jones} D.,   {Skarka} M.,  eds, ,
  Reviews in Frontiers of Modern Astrophysics; From Space Debris to Cosmology.
Cham, Switzerland: Springer Nature, pp 189--211,
  \mn@doi{10.1007/978-3-030-38509-5\_7}

\bibitem[\protect\citeauthoryear{{Pejcha} \& {Prieto}}{{Pejcha} \&
  {Prieto}}{2015a}]{2015ApJ...799..215P}
{Pejcha} O.,  {Prieto} J.~L.,  2015a, \mn@doi [\apj]
  {10.1088/0004-637X/799/2/215}, \href
  {http://adsabs.harvard.edu/abs/2015ApJ...799..215P} {799, 215}

\bibitem[\protect\citeauthoryear{{Pejcha} \& {Prieto}}{{Pejcha} \&
  {Prieto}}{2015b}]{2015ApJ...806..225P}
{Pejcha} O.,  {Prieto} J.~L.,  2015b, \mn@doi [\apj]
  {10.1088/0004-637X/806/2/225}, \href
  {http://adsabs.harvard.edu/abs/2015ApJ...806..225P} {806, 225}

\bibitem[\protect\citeauthoryear{{Perez} \& {Granger}}{{Perez} \&
  {Granger}}{2007}]{2007CSE.....9c..21P}
{Perez} F.,  {Granger} B.~E.,  2007, \mn@doi [Computing in Science and
  Engineering] {10.1109/MCSE.2007.53}, \href
  {https://ui.adsabs.harvard.edu/abs/2007CSE.....9c..21P} {9, 21}

\bibitem[\protect\citeauthoryear{{Pignata}}{{Pignata}}{2013}]{2013msao.confE.176P}
{Pignata} G.,  2013, in Massive Stars: From alpha to Omega. p.~176

\bibitem[\protect\citeauthoryear{{Popov}}{{Popov}}{1993}]{1993ApJ...414..712P}
{Popov} D.~V.,  1993, \mn@doi [\apj] {10.1086/173117}, \href
  {http://adsabs.harvard.edu/abs/1993ApJ...414..712P} {414, 712}

\bibitem[\protect\citeauthoryear{{Pumo} et~al.,}{{Pumo}
  et~al.}{2009}]{2009ApJ...705L.138P}
{Pumo} M.~L.,  et~al., 2009, \mn@doi [\apjl] {10.1088/0004-637X/705/2/L138},
  \href {https://ui.adsabs.harvard.edu/abs/2009ApJ...705L.138P} {705, L138}

\bibitem[\protect\citeauthoryear{{Pumo}, {Zampieri}, {Spiro}, {Pastorello},
  {Benetti}, {Cappellaro}, {Manic{\`o}}  \& {Turatto}}{{Pumo}
  et~al.}{2017}]{2017MNRAS.464.3013P}
{Pumo} M.~L.,  {Zampieri} L.,  {Spiro} S.,  {Pastorello} A.,  {Benetti} S.,
  {Cappellaro} E.,  {Manic{\`o}} G.,   {Turatto} M.,  2017, \mn@doi [\mnras]
  {10.1093/mnras/stw2625}, \href
  {https://ui.adsabs.harvard.edu/abs/2017MNRAS.464.3013P} {464, 3013}

\bibitem[\protect\citeauthoryear{{Rampp} \& {Janka}}{{Rampp} \&
  {Janka}}{2002}]{2002A&A...396..361R}
{Rampp} M.,  {Janka} H.~T.,  2002, \mn@doi [\aap] {10.1051/0004-6361:20021398},
  \href {https://ui.adsabs.harvard.edu/abs/2002A&A...396..361R} {396, 361}

\bibitem[\protect\citeauthoryear{{Reguitti} et~al.,}{{Reguitti}
  et~al.}{2021}]{2021MNRAS.501.1059R}
{Reguitti} A.,  et~al., 2021, \mn@doi [\mnras] {10.1093/mnras/staa3730}, \href
  {https://ui.adsabs.harvard.edu/abs/2021MNRAS.501.1059R} {501, 1059}

\bibitem[\protect\citeauthoryear{{Rembiasz}, {Obergaulinger}, {Masip},
  {P{\'e}rez-Garc{\'\i}a}, {Aloy}  \& {Albertus}}{{Rembiasz}
  et~al.}{2018}]{Rembiasz+2018}
{Rembiasz} T.,  {Obergaulinger} M.,  {Masip} M.,  {P{\'e}rez-Garc{\'\i}a}
  M.~A.,  {Aloy} M.~A.,   {Albertus} C.,  2018, \mn@doi [\prd]
  {10.1103/PhysRevD.98.103010}, \href
  {https://ui.adsabs.harvard.edu/abs/2018PhRvD..98j3010R} {98, 103010}

\bibitem[\protect\citeauthoryear{{Scheck}, {Kifonidis}, {Janka}  \&
  {M{\"u}ller}}{{Scheck} et~al.}{2006}]{2006A&A...457..963S}
{Scheck} L.,  {Kifonidis} K.,  {Janka} H.-T.,   {M{\"u}ller} E.,  2006, \mn@doi
  [\aap] {10.1051/0004-6361:20064855}, \href
  {http://adsabs.harvard.edu/abs/2006A%26A...457..963S} {457, 963}

\bibitem[\protect\citeauthoryear{{Spiro} et~al.,}{{Spiro}
  et~al.}{2014}]{2014MNRAS.439.2873S}
{Spiro} S.,  et~al., 2014, \mn@doi [\mnras] {10.1093/mnras/stu156}, \href
  {https://ui.adsabs.harvard.edu/abs/2014MNRAS.439.2873S} {439, 2873}

\bibitem[\protect\citeauthoryear{{Stockinger} et~al.,}{{Stockinger}
  et~al.}{2020}]{2020MNRAS.496.2039S}
{Stockinger} G.,  et~al., 2020, \mn@doi [\mnras] {10.1093/mnras/staa1691},
  \href {https://ui.adsabs.harvard.edu/abs/2020MNRAS.496.2039S} {496, 2039}

\bibitem[\protect\citeauthoryear{{Sukhbold}, {Ertl}, {Woosley}, {Brown}  \&
  {Janka}}{{Sukhbold} et~al.}{2016}]{2016ApJ...821...38S}
{Sukhbold} T.,  {Ertl} T.,  {Woosley} S.~E.,  {Brown} J.~M.,   {Janka} H.-T.,
  2016, \mn@doi [\apj] {10.3847/0004-637X/821/1/38}, \href
  {http://adsabs.harvard.edu/abs/2016ApJ...821...38S} {821, 38}

\bibitem[\protect\citeauthoryear{{Summa}, {Janka}, {Melson}  \&
  {Marek}}{{Summa} et~al.}{2018}]{2018ApJ...852...28S}
{Summa} A.,  {Janka} H.-T.,  {Melson} T.,   {Marek} A.,  2018, \mn@doi [\apj]
  {10.3847/1538-4357/aa9ce8}, \href
  {https://ui.adsabs.harvard.edu/abs/2018ApJ...852...28S} {852, 28}

\bibitem[\protect\citeauthoryear{{Takiwaki}, {Kotake}  \& {Suwa}}{{Takiwaki}
  et~al.}{2014}]{2014ApJ...786...83T}
{Takiwaki} T.,  {Kotake} K.,   {Suwa} Y.,  2014, \mn@doi [\apj]
  {10.1088/0004-637X/786/2/83}, \href
  {https://ui.adsabs.harvard.edu/abs/2014ApJ...786...83T} {786, 83}

\bibitem[\protect\citeauthoryear{{Tody}}{{Tody}}{1986}]{1986SPIE..627..733T}
{Tody} D.,  1986, in {Crawford} D.~L.,  ed.,  Society of Photo-Optical
  Instrumentation Engineers (SPIE) Conference Series Vol. 627, Instrumentation
  in astronomy VI. p.~733, \mn@doi{10.1117/12.968154}

\bibitem[\protect\citeauthoryear{{Tomasella} et~al.,}{{Tomasella}
  et~al.}{2018}]{2018MNRAS.475.1937T}
{Tomasella} L.,  et~al., 2018, \mn@doi [\mnras] {10.1093/mnras/stx3220}, \href
  {https://ui.adsabs.harvard.edu/abs/2018MNRAS.475.1937T} {475, 1937}

\bibitem[\protect\citeauthoryear{{Tsvetkov}, {Volnova}, {Shulga}, {Korotkiy},
  {Elmhamdi}, {Danziger}  \& {Ereshko}}{{Tsvetkov}
  et~al.}{2006}]{2006AandA...460..769T}
{Tsvetkov} D.~Y.,  {Volnova} A.~A.,  {Shulga} A.~P.,  {Korotkiy} S.~A.,
  {Elmhamdi} A.,  {Danziger} I.~J.,   {Ereshko} M.~V.,  2006, \mn@doi [\aap]
  {10.1051/0004-6361:20065704}, \href
  {https://ui.adsabs.harvard.edu/abs/2006A&A...460..769T} {460, 769}

\bibitem[\protect\citeauthoryear{{Utrobin} \& {Chugai}}{{Utrobin} \&
  {Chugai}}{2008}]{2008AandA...491..507U}
{Utrobin} V.~P.,  {Chugai} N.~N.,  2008, \mn@doi [\aap]
  {10.1051/0004-6361:200810272}, \href
  {https://ui.adsabs.harvard.edu/abs/2008A&A...491..507U} {491, 507}

\bibitem[\protect\citeauthoryear{{Utrobin}, {Wongwathanarat}, {Janka}  \&
  {M{\"u}ller}}{{Utrobin} et~al.}{2017}]{2017ApJ...846...37U}
{Utrobin} V.~P.,  {Wongwathanarat} A.,  {Janka} H.-T.,   {M{\"u}ller} E.,
  2017, \mn@doi [\apj] {10.3847/1538-4357/aa8594}, \href
  {http://adsabs.harvard.edu/abs/2017ApJ...846...37U} {846, 37}

\bibitem[\protect\citeauthoryear{{Valerin} et~al.,}{{Valerin}
  et~al.}{2022}]{2022arXiv220303988V}
{Valerin} G.,  et~al., 2022, arXiv e-prints, \href
  {https://ui.adsabs.harvard.edu/abs/2022arXiv220303988V} {p. arXiv:2203.03988}

\bibitem[\protect\citeauthoryear{{Van Dyk} et~al.,}{{Van Dyk}
  et~al.}{2012}]{2012AJ....143...19V}
{Van Dyk} S.~D.,  et~al., 2012, \mn@doi [\aj] {10.1088/0004-6256/143/1/19},
  \href {https://ui.adsabs.harvard.edu/abs/2012AJ....143...19V} {143, 19}

\bibitem[\protect\citeauthoryear{{Weaver}, {Zimmerman}  \& {Woosley}}{{Weaver}
  et~al.}{1978}]{1978ApJ...225.1021W}
{Weaver} T.~A.,  {Zimmerman} G.~B.,   {Woosley} S.~E.,  1978, \mn@doi [\apj]
  {10.1086/156569}, \href {http://adsabs.harvard.edu/abs/1978ApJ...225.1021W}
  {225, 1021}

\bibitem[\protect\citeauthoryear{{Wongwathanarat}, {M{\"u}ller}  \&
  {Janka}}{{Wongwathanarat} et~al.}{2015}]{2015A&A...577A..48W}
{Wongwathanarat} A.,  {M{\"u}ller} E.,   {Janka} H.-T.,  2015, \mn@doi [\aap]
  {10.1051/0004-6361/201425025}, \href
  {http://adsabs.harvard.edu/abs/2015A%26A...577A..48W} {577, A48}

\bibitem[\protect\citeauthoryear{{Woosley} \& {Heger}}{{Woosley} \&
  {Heger}}{2015}]{2015ApJ...810...34W}
{Woosley} S.~E.,  {Heger} A.,  2015, \mn@doi [\apj]
  {10.1088/0004-637X/810/1/34}, \href
  {http://adsabs.harvard.edu/abs/2015ApJ...810...34W} {810, 34}

\bibitem[\protect\citeauthoryear{{Yang} et~al.,}{{Yang}
  et~al.}{2021}]{2021AandA...655A..90Y}
{Yang} S.,  et~al., 2021, \mn@doi [\aap] {10.1051/0004-6361/202141244}, \href
  {https://ui.adsabs.harvard.edu/abs/2021A&A...655A..90Y} {655, A90}

\makeatother
\end{thebibliography}

\appendix
\section[Profiles of the angle-averaged explosion-model data used in Kozyreva et al. (2021) and in the present study]
{Profiles of the angle-averaged explosion-model data used in Kozyreva et al. (2021) and in the present study}
\label{appendix:append1}

\begin{figure*}
\centering
\vspace{4mm}\includegraphics[width=0.7\textwidth]{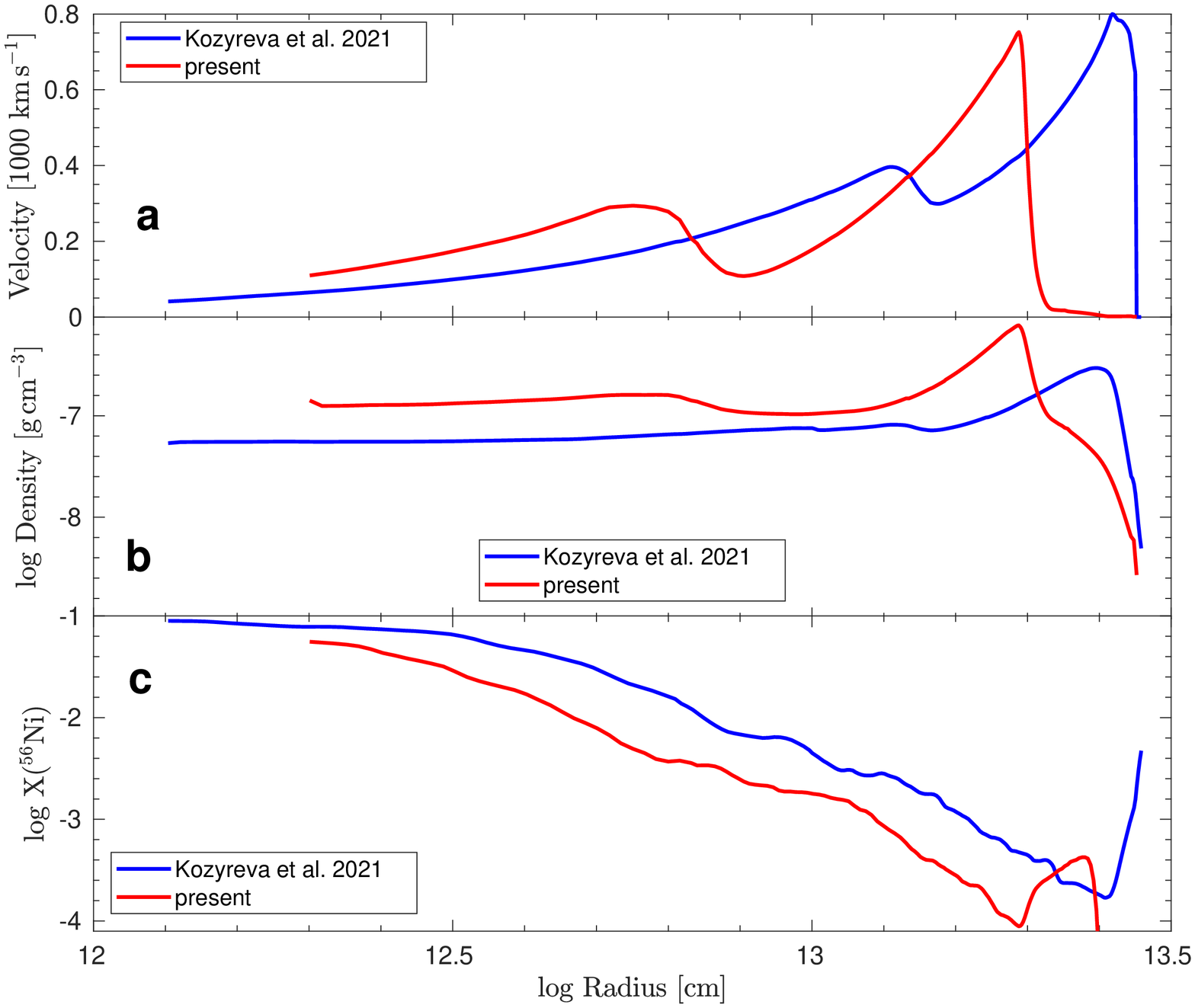}
\caption{Velocity (a), density (b) and {}$^{56}$Ni{} mass fraction (c) of the 
angle-averaged profiles used in the present study and in \citet{2021MNRAS.503..797K} along the radius coordinate.
The red lines correspond to the profiles at 1.97~days after the onset of the explosion and the blue lines to the profiles at 2.82~days.} 
\label{figure:strucR}
\end{figure*}

\begin{figure*}
\centering
\vspace{4mm}\includegraphics[width=0.7\textwidth]{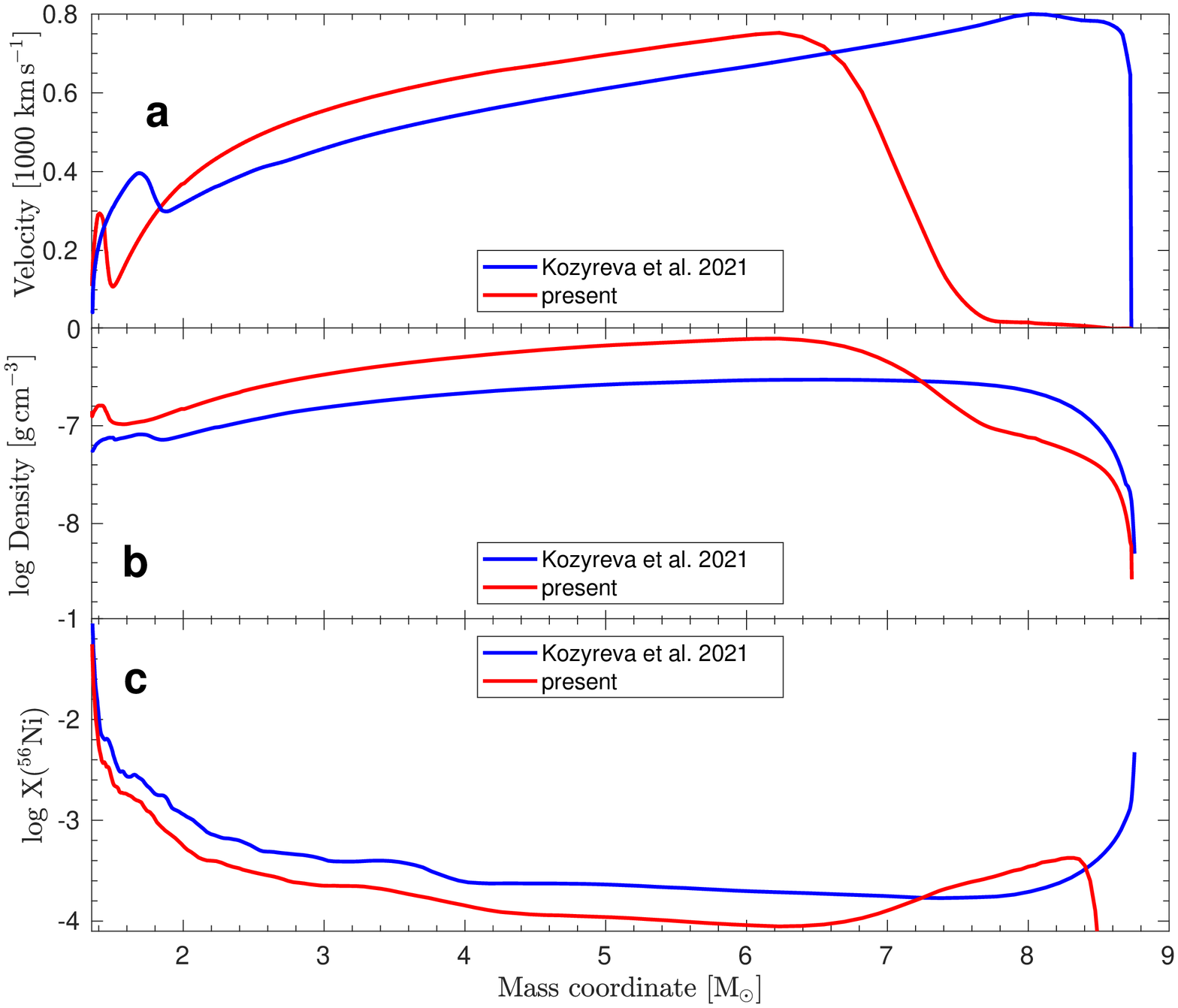}
\caption{Velocity (a), density (b) and {}$^{56}$Ni{} mass fraction (c) of the 
angle-averaged profiles used in the present study and in \citet{2021MNRAS.503..797K} along the
mass coordinate.
The red lines correspond to the profiles at 1.97~days after the onset of the explosion and the blue lines to the profiles at 2.82~days.} 
\label{figure:strucM}
\end{figure*}

We show the employed initial profiles of the angle-averaged velocity, density,
and $^{56}$Ni mass fraction versus radius (Figure~\ref{figure:strucR}) and enclosed
mass (Figure~\ref{figure:strucM}) compared to the profiles used in \citet{2021MNRAS.503..797K}.
Instead of considering model results from \citet{2020MNRAS.496.2039S} at 2.82~days after the onset of 
the explosion as in the previous work, we started the radiation-hydrodynamic light-curve modeling in the 
present study with a model profile at 1.97~days, which is about 0.2~days before the fastest part of the 
highly deformed SN shock breaks out from the progenitor surface. This ensures a somewhat cleaner modeling of
the shock-breakout phase in our 1D simulations, because this phase takes roughly 1\,day in the non-spherical
3D explosion model of \citet{2020MNRAS.496.2039S}, leading to a corresponding broadening of the structural
features near the stellar surface.

Besides a somewhat smaller shock radius, the earlier initial profile of the present study displays
an outward going wave around a radius of $6\times 10^{12}$\,cm and enclosed mass of 1.4\,M$_\odot$  visible as velocity bump 
in the red lines. This feature is connected to the reflection of the reverse shock near the stellar 
center and has moved to a larger radius and enclosed mass in the later simulation output ($\sim$\,$1.2\times 10^{13}$\,cm and $\sim$1.7\,M$_\odot$, respectively). The extreme deformation of the pre-breakout shock in the
3D explosion model at 1.97~days is clearly obvious from the fact that the corresponding velocity jump in the
angle-averaged profile is smeared over a mass interval between 6.6\,M$_\odot$ and 8.6\,M$_\odot$. The local
maximum in the profile of the $^{56}$Ni mass fraction at radii $>$\,$2\times 10^{13}$\,cm (enclosed mass 
between $\sim$6.6\,M$_\odot$ and $\sim$8.5\,M$_\odot$) is connected to the biggest low-mass high-velocity 
nickel-rich plume prominently visible in Figure~\ref{figure:asymmetry}.

\section[Angle-averaged and radial-direction dependent profiles extracted from the 3D supernova explosion simulation]
{Angle-averaged and radial-direction dependent profiles extracted from the 3D supernova explosion simulation}
\label{appendix:append2}

\begin{figure*}
\centering
\vspace{5mm}
\hspace{-5mm}\includegraphics[width=0.5\textwidth]{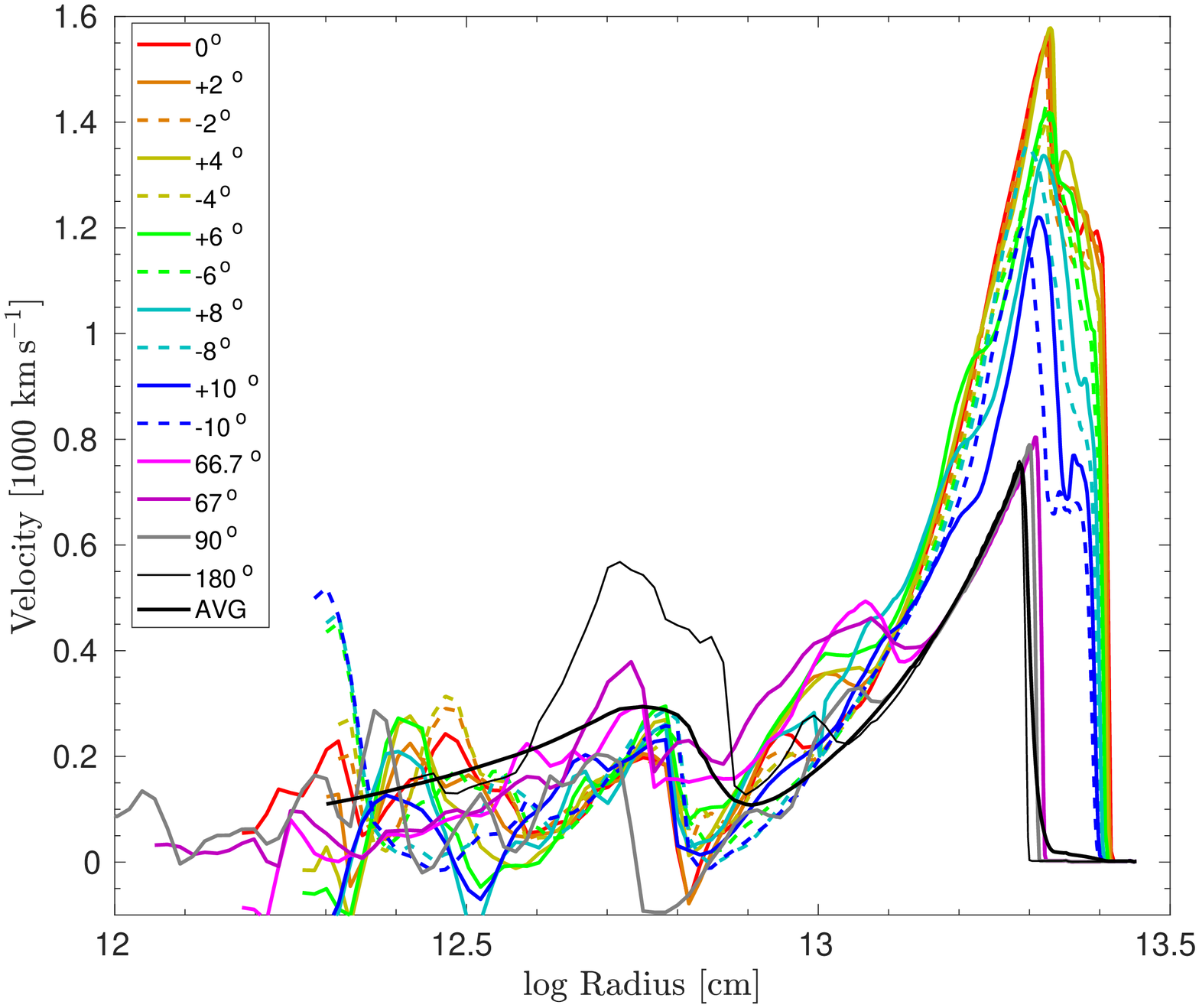}\hspace{5mm}
\includegraphics[width=0.49\textwidth]{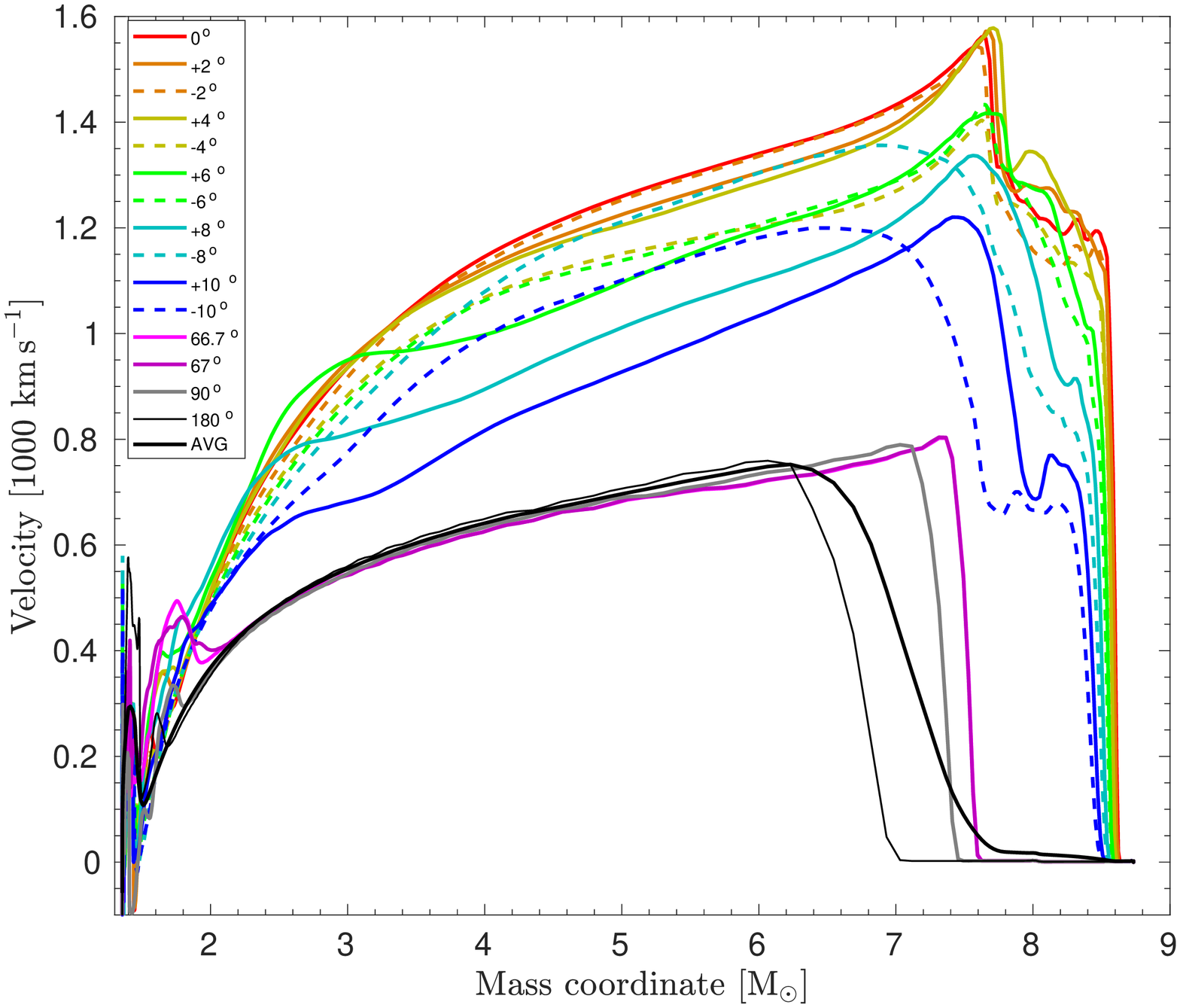}
\caption{Velocity of the angle-averaged profile (``AVG'') and different radial
directions of the model s9.0 used in the present study along the radius
coordinate (left) and the mass coordinate (right).
} 
\label{figure:raysVelo}
\end{figure*}

\begin{figure*}
\centering
\vspace{5mm}
\hspace{-5mm}\includegraphics[width=0.5\textwidth]{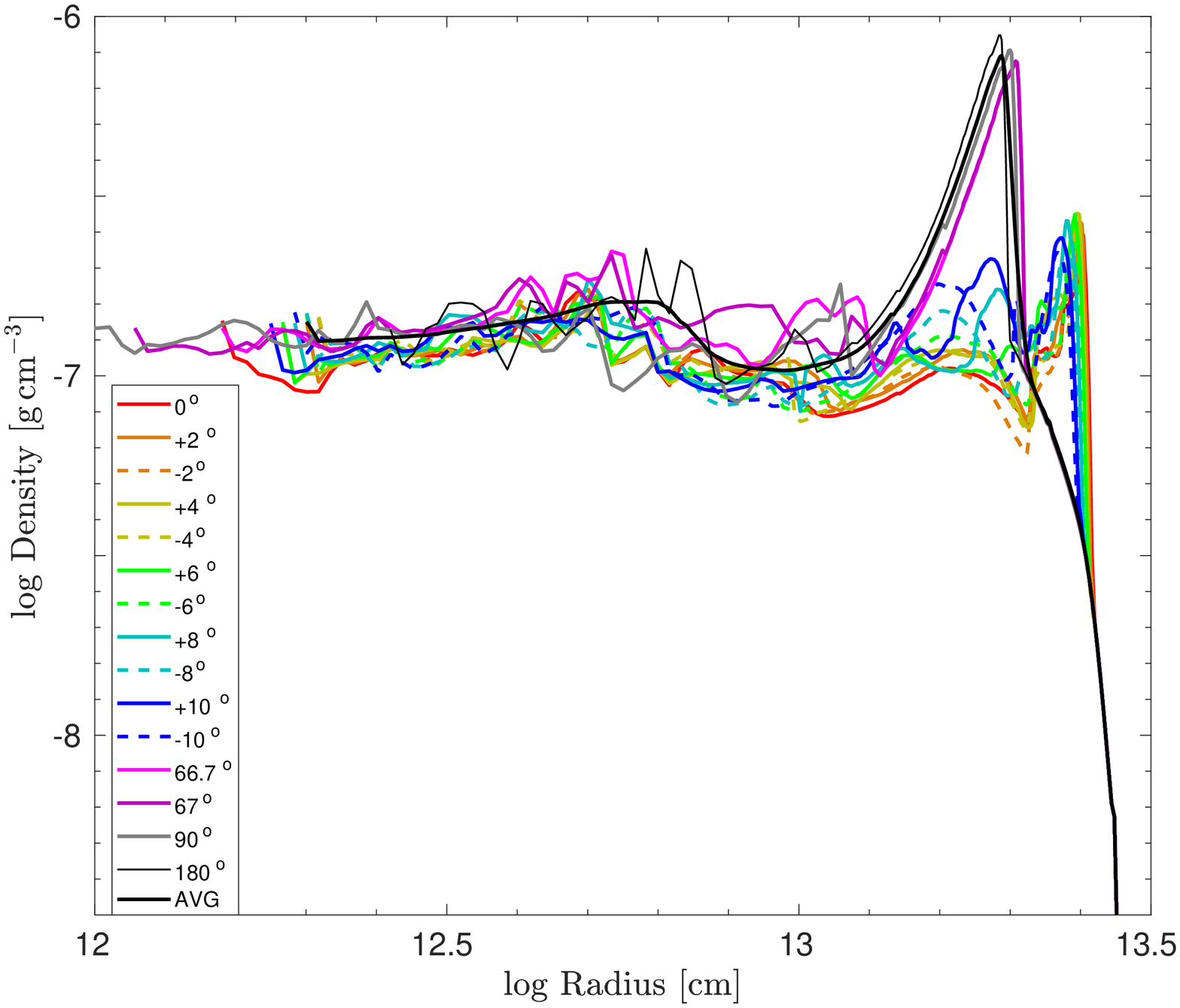}\hspace{5mm}
\includegraphics[width=0.49\textwidth]{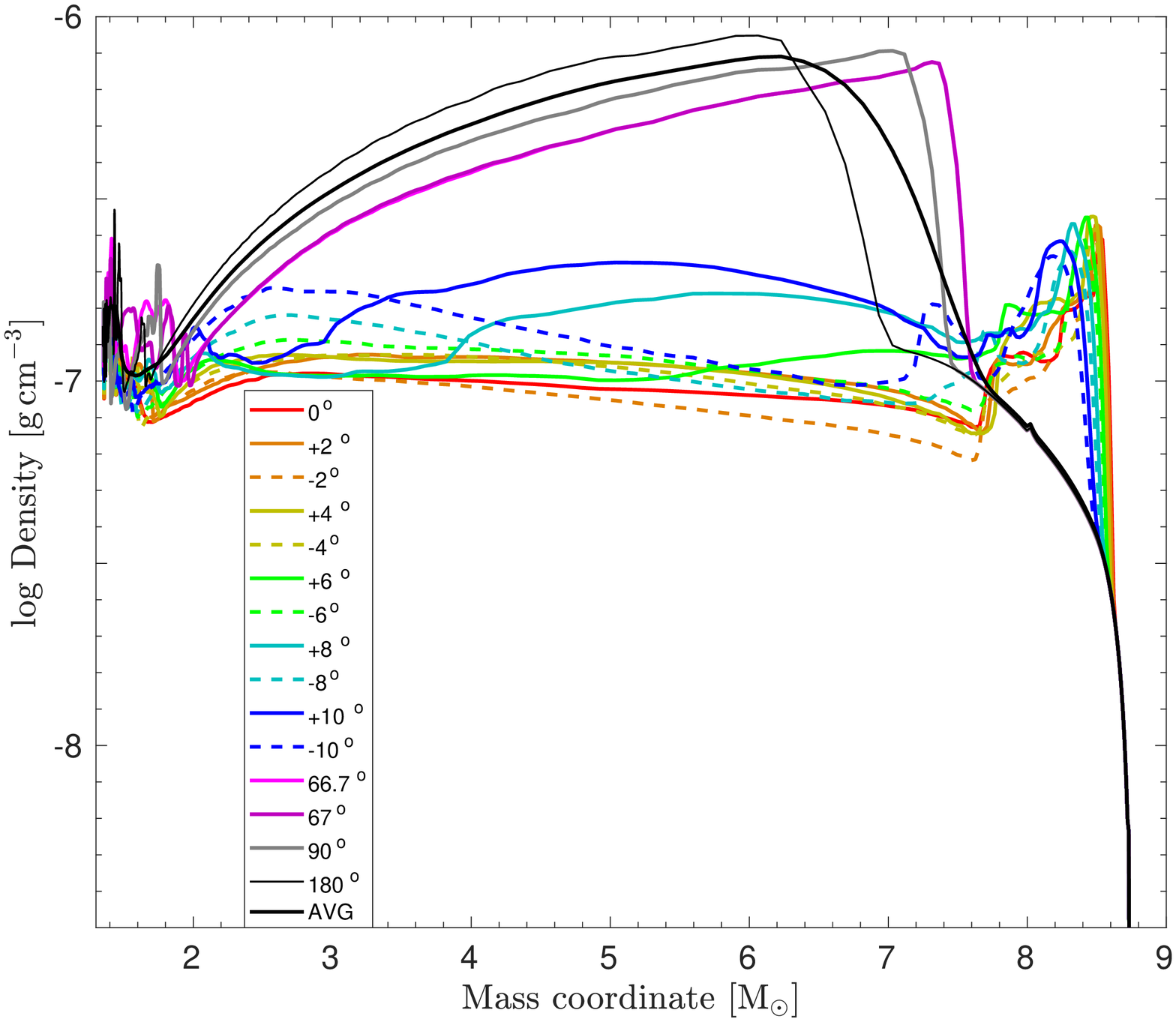}
\caption{Density of the angle-averaged profile (``AVG'') and different radial
directions of the model s9.0 used in the present study along the radius coordinate
(left) and the mass coordinate (right).
} 
\label{figure:raysRho}
\end{figure*}

\begin{figure*}
\centering
\hspace{-5mm}\includegraphics[width=0.5\textwidth]{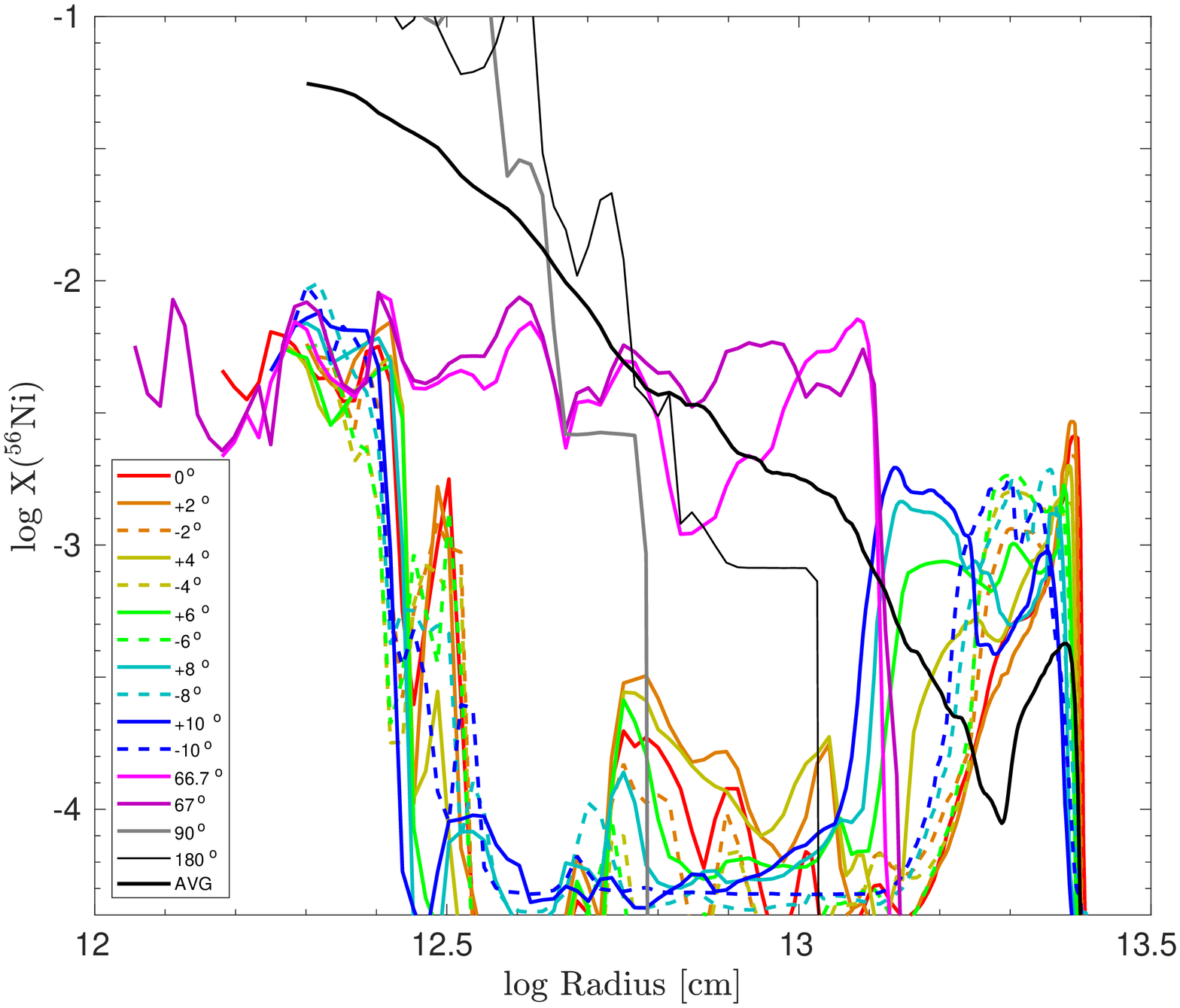}\hspace{5mm}
\includegraphics[width=0.49\textwidth]{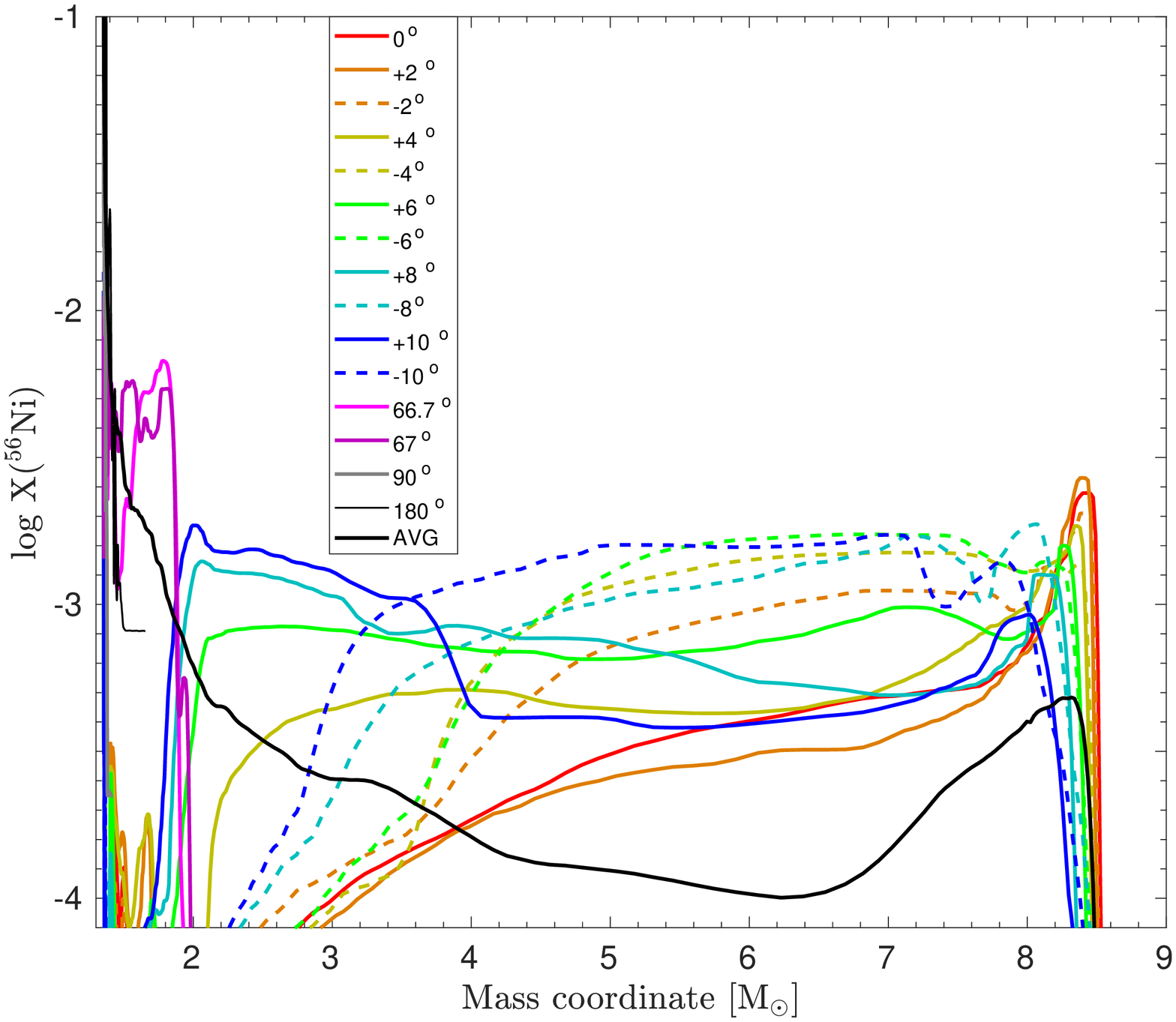}
\caption{{}$^{56}$Ni{} mass fraction of the angle-averaged profile (``AVG'') and different radial
directions of the model s9.0 used in the present study along the radius
coordinate (left) and the mass coordinate (right).
} 
\label{figure:raysNi56}
\end{figure*}

Figure~\ref{figure:raysVelo}, Figure~\ref{figure:raysRho}, and Figure~\ref{figure:raysNi56}
display the profiles of velocity, density, and {}$^{56}$Ni{} 
mass fraction, respectively, of all selected radial directions of model s9.0 that were 
used as initial data for our radiation-hydrodynamics simulations in the present
study. At the given time of 1.97~days after core collapse
the shock wave in the fastest moving parts of the SN ejecta is within about $3\times 10^{12}$\,cm from the stellar surface
and breaks out within $\sim$0.2~days. It corresponds to the great, low-mass nickel-rich
plume subtending the directions from $0^\circ$ to about $10^\circ$ in Figure~\ref{figure:asymmetry}. The shock wave in the slowest parts of the ejecta
is still about $10^{13}$\,cm away from the stellar edge and will reach the surface
roughly 1~day after the fastest regions of the shock. In the angle-averaged velocity
profile (bold black solid line) the fast-moving $^{56}$Ni plume is hardly visible because of the
small angular extension of this feature.

Similarly, the biggest plume forms a prominent cusp in the density profile around
$\sim$\,$2.5\times 10^{13}$\,cm (mass coordinate around 8.4\,M$_\odot$ for directions between
0$^\circ$ and 10$^\circ$, which moves well ahead of the main shock and the main ejecta shell with its density maximum
at about $2\times 10^{13}$\,cm. This main ejecta shell appears as a broad dome from $\sim$2\,M$_\odot$ to
$\sim$7.5\,M$_\odot$ in the density profile versus enclosed mass (Figure~\ref{figure:raysRho}).
Again, the angle-averaged profile (bold black solid line) does not carry any evidence of 
the fastest, most elongated nickel plume because of the small amount of mass in this structure. 
The lower-amplitude density and velocity fluctuations at small radii and enclosed masses
correspond to waves and density inhomogeneities in the inner, slow ejecta.

In the $^{56}$Ni distribution (Figure~\ref{figure:raysNi56}) high mass fractions 
signal the existence of a more spherical bulk mass of slower nickel that extends up to 
$\sim$10$^{13}$\,cm and has low 
expansion velocities ($\lesssim$500\,km\,s$^{-1}$; see lines for 90$^\circ$ and 180$^\circ$
directions and also Figure~\ref{figure:asymmetry}). Moreover, besides the biggest nickel-rich
plume (directions 0$^\circ$ to $\sim$10$^\circ$) high $^{56}$Ni concentrations can be 
found in secondary plumes reaching up to around $1.3\times 10^{13}$\,cm and thus overlapping with the 
innermost parts of the dense main ejecta shell (66.7$^\circ$ and 67$^\circ$ directions).

\section[Broad-band light curves and colours for all radial directions in the present study]
{Broad-band light curves and colours for all radial directions in the present study}
\label{appendix:append3}

\begin{figure*}
\centering
\vspace{5mm}
\hspace{-5mm}\includegraphics[width=0.5\textwidth]{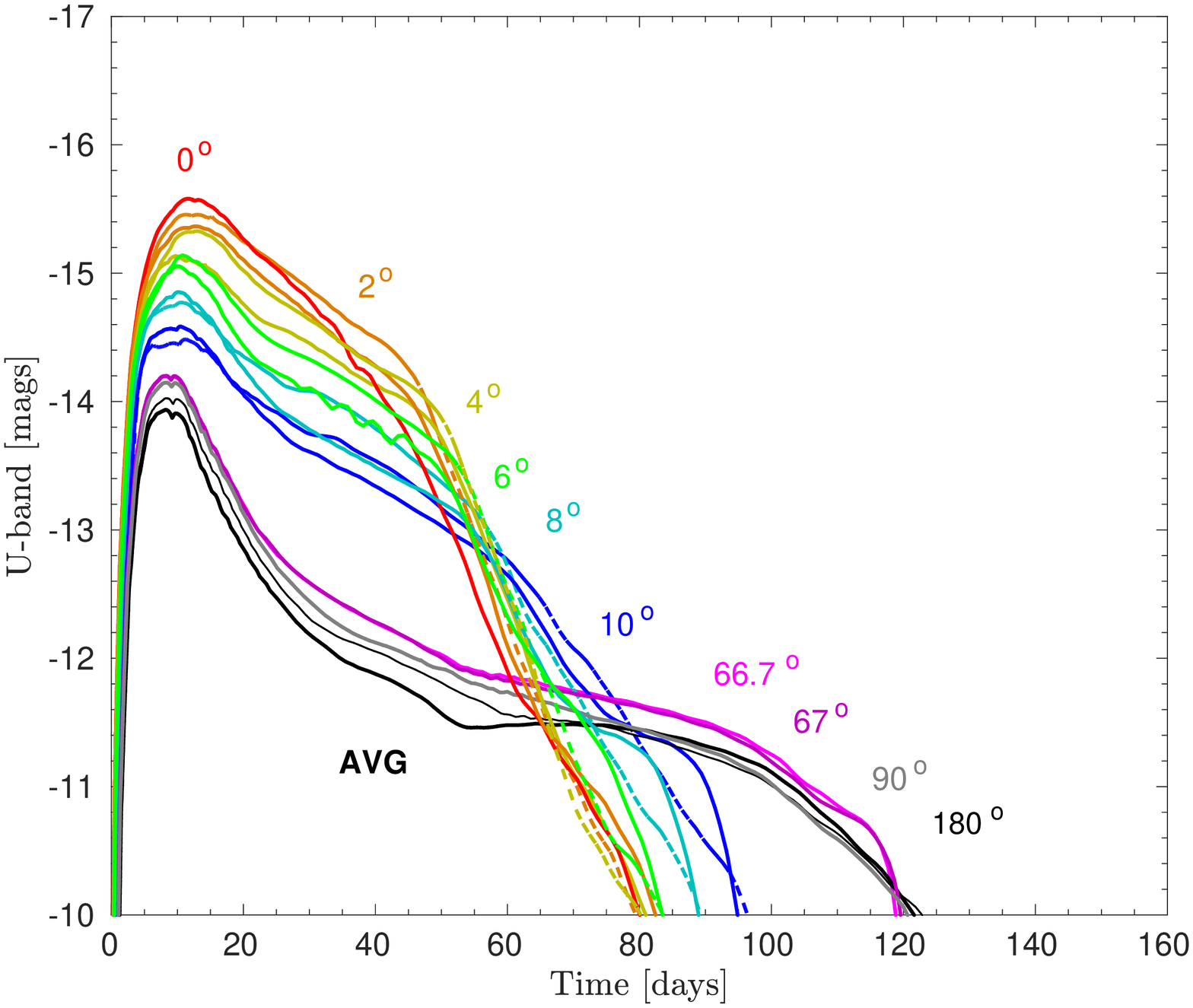}\hspace{4mm}
\includegraphics[width=0.47\textwidth]{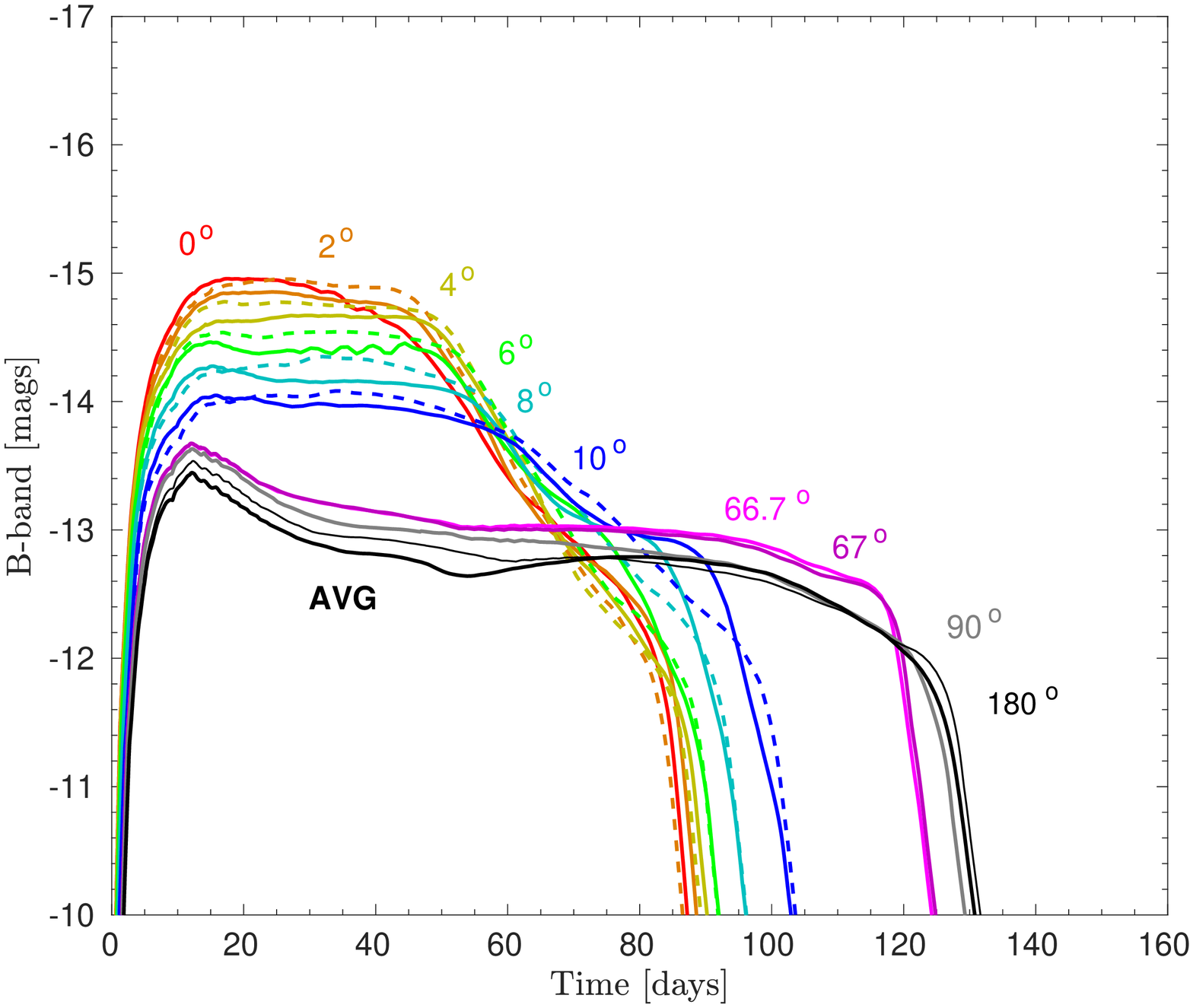}\\
\vspace{5mm}
\hspace{-5mm}\includegraphics[width=0.5\textwidth]{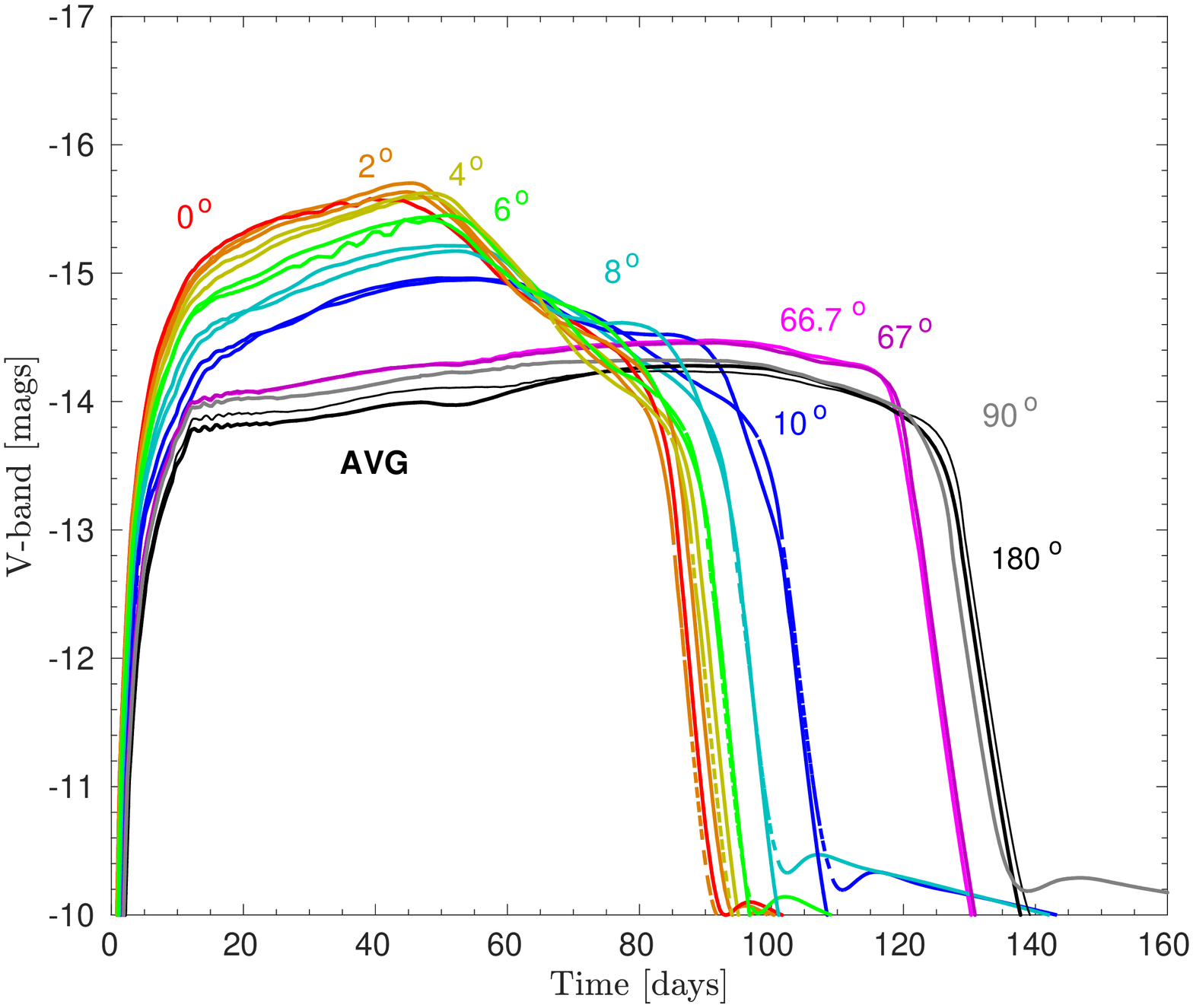}\hspace{4mm}
\includegraphics[width=0.47\textwidth]{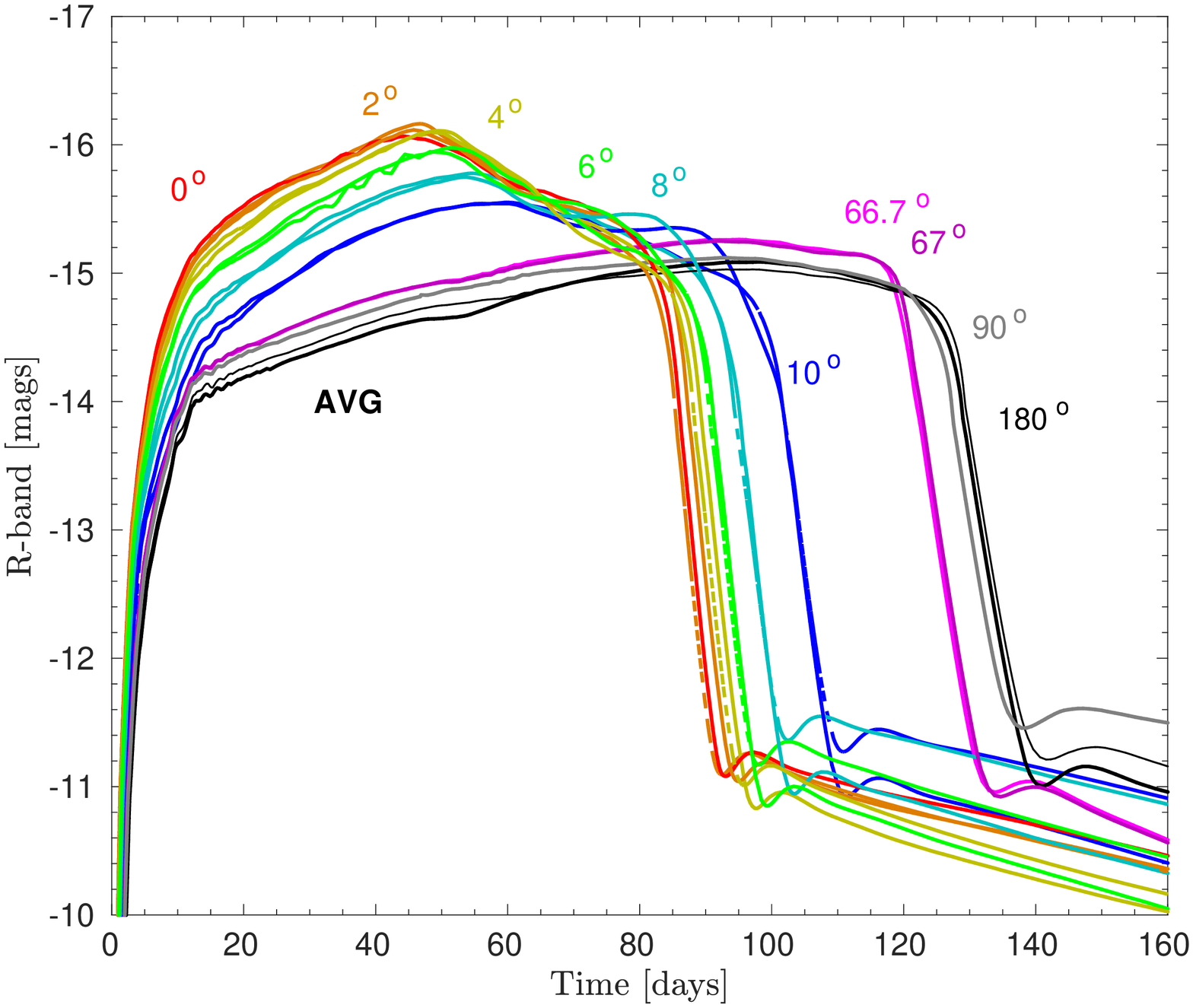}
\caption{$U\!BV\!R$ broad-band light curves for the angle-averaged profile (``AVG'')
and different radial directions of the 3D model s9.0.
The solid and dashed curves indicate the ``$+$'' and ``$-$'' directions for the cases of 2$^\circ$--10$^\circ$, as introduced in Table~\ref{table:rays}.}
\label{figure:bands}
\end{figure*}

We present the $U\!BV\!R$ broad-band magnitudes for all radial directions
considered in our study and the angle-averaged profile of model s9.0 in
Figure~\ref{figure:bands}.
The broad-band light curves behave similarly to the bolometric light curves and reflect the properties of the profiles for the different considered radial directions. In the radial directions aligned with the main nickel-rich plume, i.e., for 0$^\circ${}, 2$^\circ${}, 4$^\circ${}, 6$^\circ${}, 8$^\circ${}, and 10$^\circ${}, radioactive nickel is distributed such that a relatively large fraction of it is located close to the stellar surface because of the deep penetration of the plume into the hydrogen envelope (see Figure~\ref{figure:raysNi56}). Moreover, the density distribution in these radial directions indicates lower density along the main plume than in its surroundings (see Figure~\ref{figure:raysRho}). Therefore the $4\pi$-equivalent of the total mass that is relevant when the data are sphericized and mapped into the 1D treatment with \verb|STELLA| is relatively low compared to the radial directions outside of the main plume and the angle-averaged profile (see Table~\ref{table:rays}). For example, the effective ejecta mass for the 0$^\circ${} direction is only 4.4~\Msun{}. Note, however, that the stellar radii for all radial directions are the same. Since the bulk velocity of the matter in the main plume is higher than that along the radial directions outside of it (see Figure~\ref{figure:raysVelo}), the effective explosion energy after spherisation reaches $\sim$0.13~foe, whereas the $4\pi$-equivalent values of the explosion energies for the $66.7^\circ\!-\!180^\circ${} directions are only 0.07--0.08~foe. 

Consequently, in the radial directions away from the high-entropy plume the light curve shapes of all broad bands are comparable to those of characteristic SNe~IIP, even though they reach lower luminosities because of the lower explosion energies than canonical type~II SNe. In contrast, all radial directions along the main plume represent high-energy explosions with low-mass ejecta. Their light curves are shorter and possess higher luminosities, and their $U$-band magnitudes are higher than that of the angle-averaged case, because the plume contains hotter material. We note that the amount of {}$^{56}$Ni{} is scaled to 0.003~\Msun{}, for which reason the light curves for all radial directions including those inside the main plume are not powered by {}$^{56}$Ni{} decay but by recombination. The light curves for all of the radial directions within $0^\circ\!-\!10^\circ$ exhibit a knee around day~40--50, when 
recombination sets in in the ejecta. A similar behaviour is observed in the ECSN model e8.8 published by \citet{2021MNRAS.503..797K}; especially the $U$-band magnitude of this model exhibits a gentle rise of the plateau until about day 50. In the case of the ECSN model this long period of time until recombination starts is explained by the large progenitor radius. In the cases of the radial directions along the main plume of our s9.0 model, the reason for the delayed recombination is the relatively hotter material in this plume.

\begin{figure*}
\vspace{3mm}
\centering
\hspace{-6mm}\includegraphics[width=0.5\textwidth]{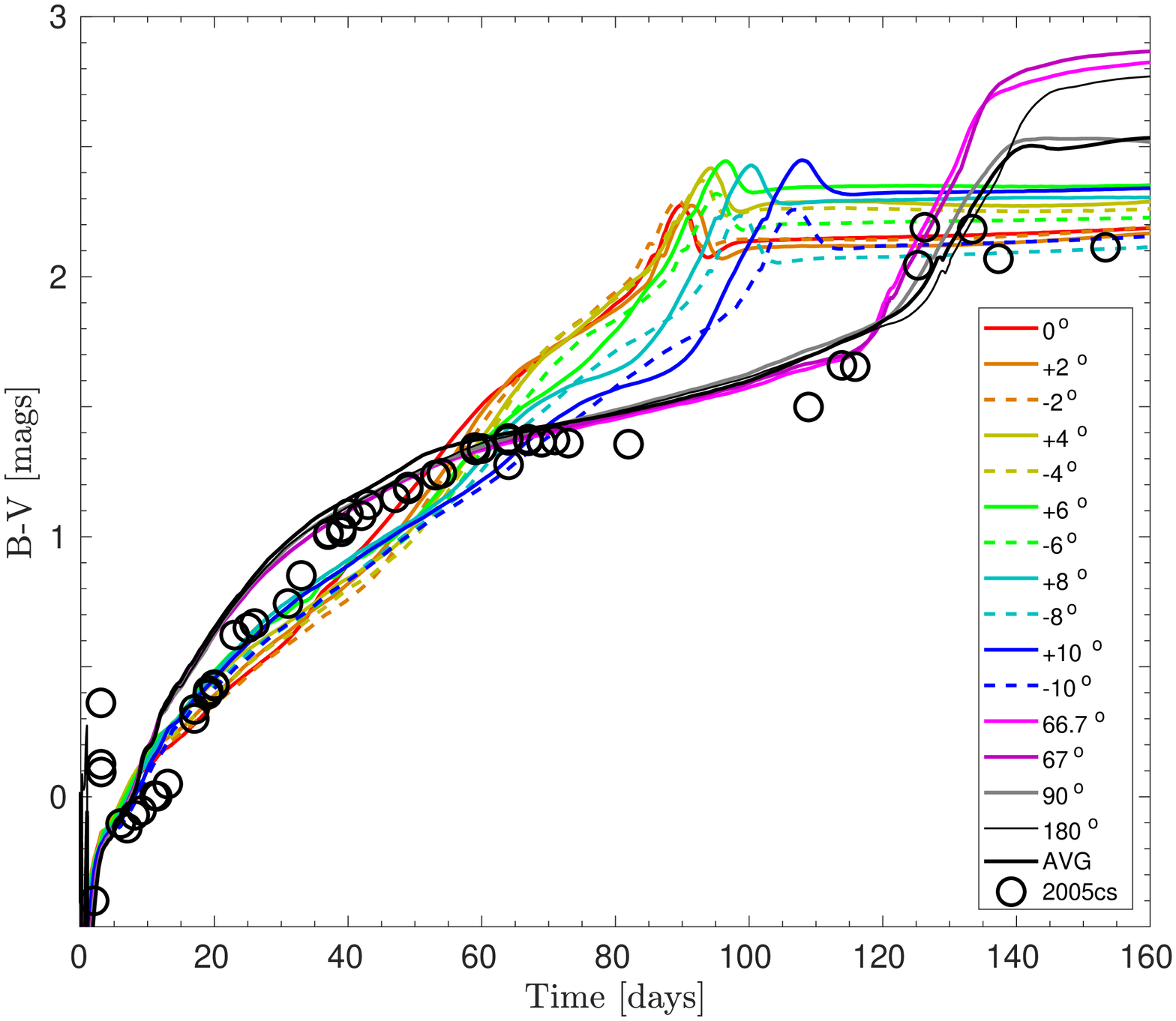}\hspace{5mm}
\includegraphics[width=0.5\textwidth]{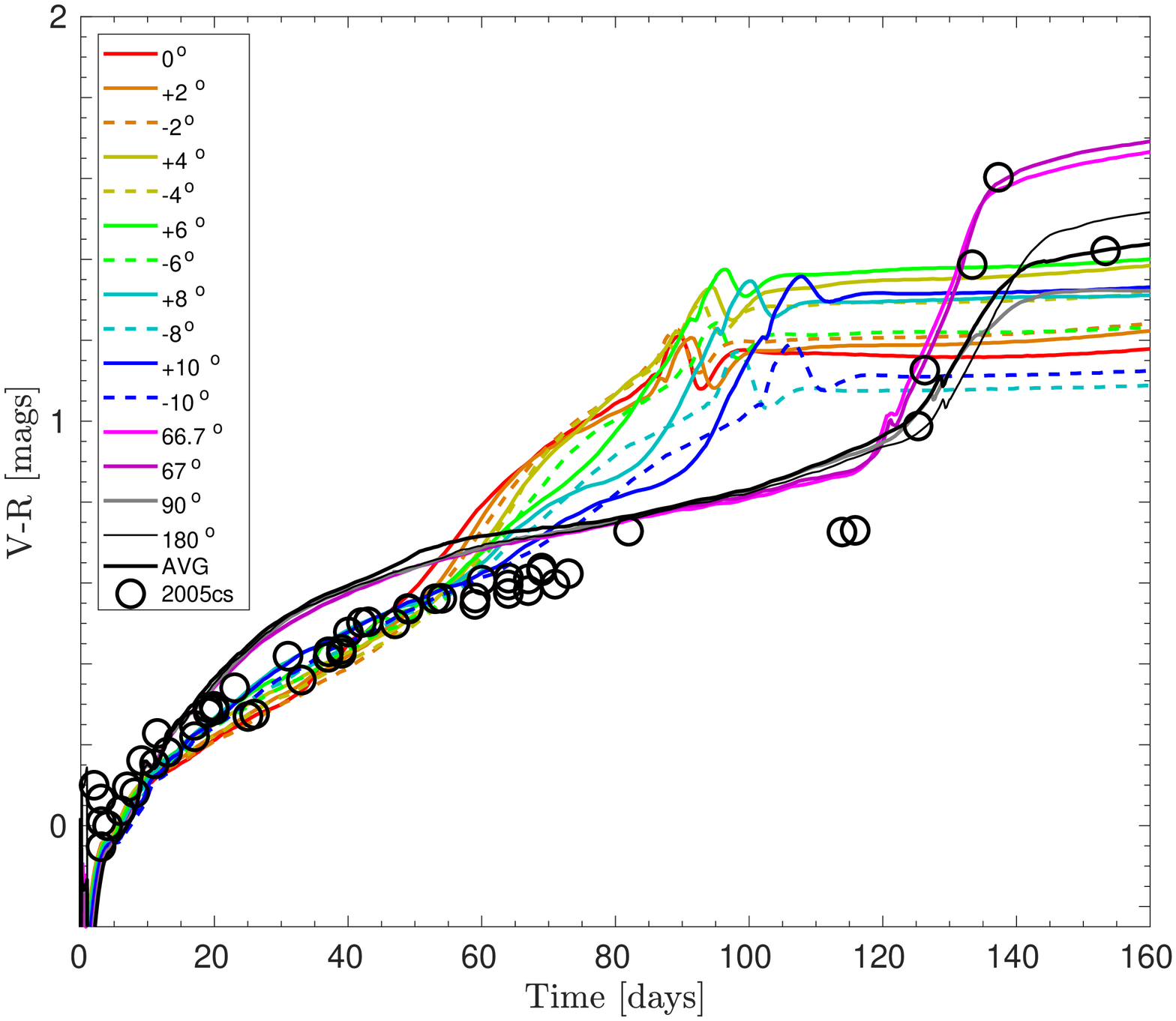}
\caption{$B\!-\!V$ and $V\!-\!R$ colour evolution for all considered radial directions in comparison to SN~2005cs. The observational data for SN~2005cs (circles) are taken from  \citet{2009MNRAS.394.2266P}.}
\label{figure:colours}
\end{figure*}

In Figure~\ref{figure:colours}, we present the $B\!-\!V$ and $V\!-\!R$ colour evolution for all considered radial directions of our 3D model s9.0 with the data for SN~2005cs superposed. \citet{2019MNRAS.483.1211K} pointed out that the $B\!-\!V$ colour may serve as a diagnostic parameter for the macroscopic mixing processes that take place in the SN ejecta after the passage of the shock wave [the interested reader is referred to a discussion of this hydrodynamical phenomenon in recent 3D explosion models of type IIP SNe by \citet{2015A&A...577A..48W} and \citet{2017ApJ...846...37U}].
The colours indicate the temperature of the SED and indirectly reflect the degree of absorption in the SN ejecta. Since iron-group elements are the main contributors to the line opacity, their distribution determines how red or blue the SN spectra are. Therefore, the colours are blue, i.e., $B\!-\!V$ is close to 0, during the early time when the photosphere is still far from the iron-rich interior of the SN ejecta. Later, at the end of the plateau, $B\!-\!V$ approaches 2--3~mag, indicating significant redistribution of the flux between the spectral bands caused by the higher fraction of iron-group elements in the inner part of the ejecta. 

During the first 50~days the $B\!-\!V$ and $V\!-\!R$ colours for the radial directions between $0^\circ$ and $10^\circ$ remain slightly ($0.2-0.3$~mags) bluer than those for the directions outside of the main plume, because the ejecta within the plume are less massive and carry a higher energy (see Table~\ref{table:rays}), for which reason the resulting radiation is hotter.

The comparison to SN~2005cs reveals that our explosion model can nicely reproduce the overall evolution of the observational data. In the time interval between about day 10 and day 30--50, there is a slight tendency that the observations are matched better by the colours for the radial directions within $0^\circ\!-\!10^\circ$, whereas at later times the observations closely follow the colour evolution for the angle-averaged case as well as for the radial directions at $66.7^\circ$, $67^\circ$, $90^\circ$, and $180^\circ$. These findings are in line with the trends we witnessed for our synthetic bolometric light curves compared to SN~2005cs in Figure~\ref{figure:bol}, and for our theoretical photospheric velocities in comparison to the data of SN~2005cs in Figure~\ref{figure:uph}.

%


\bsp	
\label{lastpage}
\end{document}